\documentclass[onecolumn]{elsarticle}%
\usepackage{amsmath}
\usepackage{amssymb}%
\usepackage{amsfonts}%
\usepackage{graphicx}
\journal{Journal}
\begin{document}
\begin{frontmatter}
\title{Condensation of a tetrahedra rigid-body libration mode in HoBaCo$_{4}$O$_{7}$ : the origin of phase transition at 355 K}
\author[a,b]{A. I. Rykov}
\ead{rykov@woody.ocn.ne.jp}
\author[b]{Y. Ueda}
\author[b]{M. Isobe}
\author[c]{N.  Nakayama}
\author[d]{Yu.T. Pavlyukhin}
\author[d]{S. A. Petrov}
\author[a,e]{A.N. Shmakov}
\author[a,e]{V.N. Kriventsov}
\author[f]{A.N. Vasiliev}
\address[a]{Siberian Synchrotron Radiation Center, Budker INP, Lavrentieva 11, Novosibirsk, 630090, Russia}
\address[b]{Institute for Solid State Physics, University of Tokyo, 5-1-5 Kashiwanoha, Kashiwa, Chiba 277-8581, Japan}
\address[c]{Department of Advanced Materials Science and Engineering, Yamaguchi University, Ube 755-8611, Japan}
\address[d]{Institute of Solid State Chemistry MC, Siberian Branch RAS, Novosibirsk 630128, Russia}
\address[e]{Boreskov Institute of Catalysis, Siberian Branch RAS, Lavrentieva 5, Novosibirsk 630090, Russia}
\address[f]{Low Temperature Physics Department, Moscow State University, Moscow 119991, Russia}
\begin{abstract}
Rietveld profiles, M\"ossbauer spectra and x-ray absorption fine structure (XAFS)
were analyzed through the structural phase transition at $T_s= 355$ K in HoBaCo$_4$O$_7$.
Excess of the oxygen content over "O$_7$" was avoided via annealing the samples in argon
flow at $600\,^{\circ}\mathrm{C}$. Space groups (S.G.) $Pbn2_1c$ and $P6_3mc$ were used
to refine the structure parameters in the low- and high-temperature phases, respectively.
Additionally, the $Cmc2_1$ symmetry was considered as a concurrent model of structure
of the low-temperature phase.  In the high-temperature phase, severe anisotropy of thermal
motion of the major part of the oxygen atoms was observed. This anisotropic motion turns
to be quenched as the sample is cooled below $T_s$.  The variation of quadrupole splitting
$\varepsilon$ near $T_s$ is not similar to a steplike anomaly frequently seen at the
charge-ordering transition.  We observe instead a dip-like anomaly of the average
$\varepsilon$ near $T_s$. Narrow distribution of the electric field gradient (EFG) over different
cobalt sites is observed and explained on the basis of point-charge model.  XAFS spectra
show no evidence of significant difference between YBaCo$_4$O$_7$ ($T>T_s$) and
HoBaCo$_4$O$_7$ ($T<T_s$). The origin of the transition at $T_s$ is ascribed to the
condensation of the libration phonon mode associated with the rigid-body rotational
movements of the starlike tetrahedral units, the building blocks of kagom\'e network.
It is shown that the condensation of the libration mode is not compatible with translation
symmetry for the hexagonal S.G., but compatible for the orthorhombic S.G. The orthorhombic
lattice parameters and EFG components ($V_{xx}, V_{yy}, V_{zz}$) vary smoothly with temperature
at approaching $T_s$  and closely follow each other.
\end{abstract}
\begin{keyword}
cobaltite \sep phase transition  \sep HoBaCo$_4$O$_7$ \sep libration mode \sep rigid-unit mode \sep M\"ossbauer spectroscopy
\end{keyword}
\end{frontmatter}

\section*{Introduction}

The cobaltites REBaCo$_{4}$O$_{7}$ have displayed a structural phase
transition manifestative through stepwise changes of physical properties.
Peak-like anomalies of specific heat were observed for RE=Lu, Yb, Tm, Er in
the range of transition temperatures $T_{\text{s}}$ between 160 K (Lu) and 280
K (Er)\cite{Nakayama, Markina}. Magnetization \cite{Nakayama,Caignaert},
Seebek coefficient\cite{Maignan} and resistivity\cite{Caignaert, Maignan} drop
abruptly as temperature increases through the RE-specific value of
$T_{\text{s}}$. Because of the mixed valence of Co (averaged valence "+2.25")
there may occur different distributions of charges in the low- and
high-temperature phases. Therefore, the transition was conjectured to be
driven by the charge ordering in the Co subsystem\cite{Nakayama}.

REBaCo$_{4}$O$_{7}$ belongs to the family of swedenborgite, whose structure
was solved by Pauling et al\cite{swed} in 1935. The swedenborgite SbNaBe$_{4}%
$O$_{7}$ was described as a net built up of starlike Be$_{4}$O$_{13}$
clusters\cite{Huminicki}, and similar Co$_{4}$O$_{13}$ stars can be found in
REBaCo$_{4}$O$_{7}$. Two nonequivalent sites of Co exist in high-temperature
hexagonal phase \cite{VA,Huq}, but four Co sites in low-temperature
orthorhombic phase\cite{Huq}. Clearly, some redistribution of charge should be
associated with changing the symmetry. It remains under debate, however,
whether or not the transition itself is driven by the charge redistribution.

Both phases consist of regular 1:1 stacking of kagom\'{e} and triangular
layers of CoO$_{4}$ tetrahedra (Fig. 1). In each of the phases, there are 75\%
of Co sites in kagom\'{e} layers and 25\% of Co sites in triangular layers.
Each Co$_{4}$O$_{13}$ star thus involves three tetrahedra in K-layer and one
in T-layer. We found relative arrangement of these stars unchanged, however
their exact geometry somewhat changes through $T_{\text{s}}$.

From analyzing the bond valence sums in both the low- and high-temperature
phases Huq et al\cite{Huq} derived the preference of the Co site in the
triangular layers for the Co$^{3+}$ ion at $T<T_{\text{s}}$, but, contrarily,
for the Co$^{2+}$ ion at $T>T_{\text{s}}$. On the other hand, it was argued
recently \cite{Schweika} that the Co sites in triangular layers are generally
exhibiting shorter Co-O distances than the sites in kagom\'{e} layers,
irrespectively of the transition at $T_{\text{s}}$; the Co$^{3+}$ ions should
always prefer the site in triangular layers.%

%TCIMACRO{\FRAME{ftbpFU}{4.093in}{2.4556in}{0pt}{\Qcb{Perspective view of the
%quadruple unit cell in the low-temperature phase of ReBaCo$_{4}$O$_{7}$.
%Kagom\'{e} and triangular layers of CoO$_{4}$ tetrahedra are marked by "K" and
%"T". The quadruple unit cell of HoBaCo$_{4}$O$_{7}$ is shown with atomic
%coordinates refined in S.G. $Pbn2_{1}$. }}{\Qlb{f1}}{fig_1.eps}%
%width 4.093in;  height 2.4556in;  depth 0pt;  original-width 5.6889in;
%original-height 3.3984in;  cropleft "0";  croptop "1";  cropright "1";
%cropbottom "0";  filename 'Fig_1.EPS';file-properties "XNPEU";}}}%
%BeginExpansion
\begin{figure}
[ptb]
\begin{center}
\includegraphics[
natheight=3.398400in,
natwidth=5.688900in,
height=2.4556in,
width=4.093in
]%
{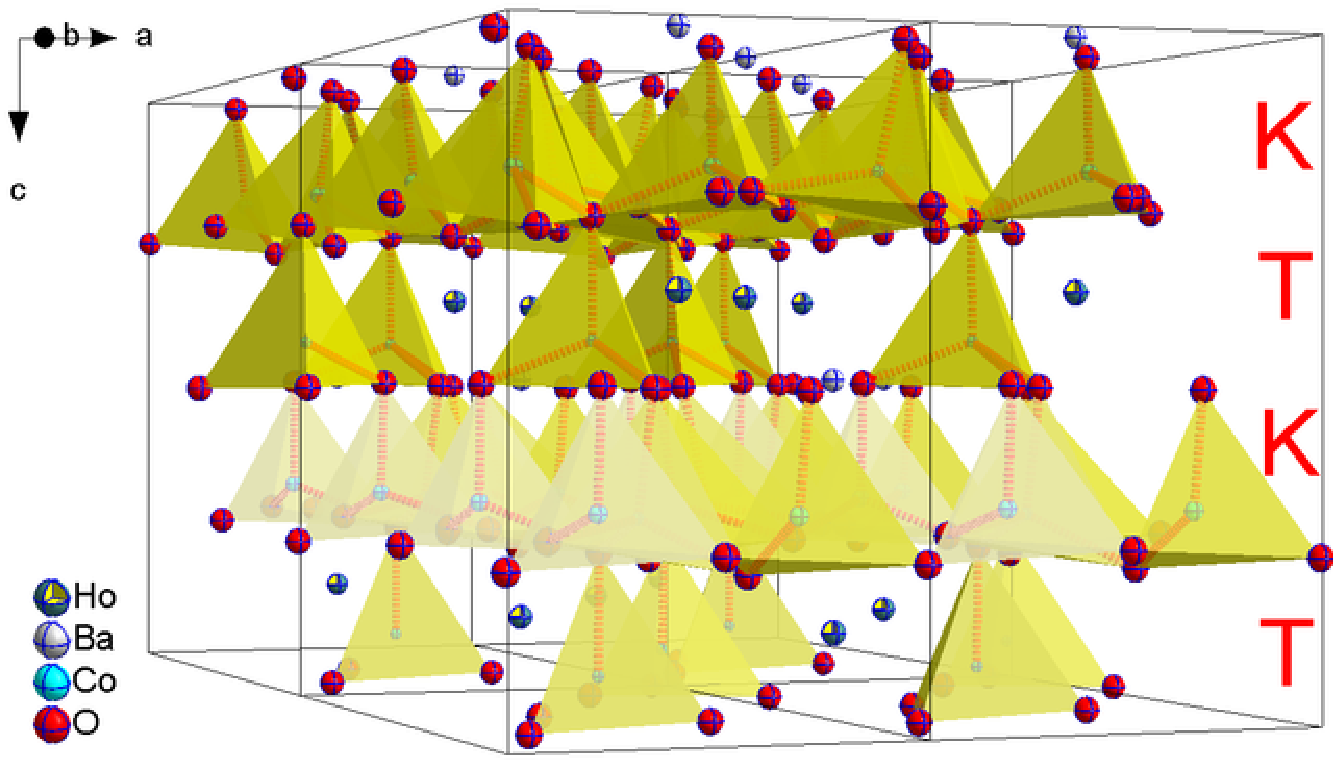}%
\caption{Perspective view of the quadruple unit cell in the low-temperature
phase of ReBaCo$_{4}$O$_{7}$. Kagom\'{e} and triangular layers of CoO$_{4}$
tetrahedra are marked by "K" and "T". The quadruple unit cell of HoBaCo$_{4}%
$O$_{7}$ is shown with atomic coordinates refined in S.G. $Pbn2_{1}$. }%
\label{f1}%
\end{center}
\end{figure}
%EndExpansion

In this work, we investigate the phase transition in HoBaCo$_{4}$O$_{7}$ using
the synchrotron diffractometry, M\"{o}ssbauer spectroscopy, and XAFS. It was
verified that the substitution of 2.5\% of Co with Fe did not introduce a
significant change of $T_{\text{s}}\simeq355$ K. We observed only very small
changes in M\"{o}ssbauer spectra through $T_{\text{s}}$ and we argue that no
charge ordering takes place at the phase transition. Different symmetry groups
were tested to fit the phases below $T_{\text{s}}$ ($Cmc2_{1}$ and $Pbn2_{1}$)
and above $T_{\text{s}}$ ($P6_{3}mc$ and $P31c$). From the results of our
structural study the origin of the phase transition is understood to be
unrelated to charge redistribution. We consider instead the double-well
potential models\cite{Dove}, in which the transition can be explained as the
condensation of a libration phonon mode associated with the rigid-body motion
of the tetrahedra in kagom\'{e} layers around the center of the abovementioned
Co$_{4}$O$_{13}$ stars.

\section{Experimental}

The samples of HoBaCo$_{4}$O$_{7}$ and HoBaCo$_{3.9}$Fe$_{0.1}$O$_{7}$ were
prepared from Ho$_{2}$O$_{3}$ , BaCO$_{3}$, Co$_{3}$O$_{4}$ and $^{57}$%
Fe$_{2}$O$_{3}$ powders using standard ceramic synthesis technology at 1100
$^{o}$C. One sample of YBaCo$_{4}$O$_{7}$ for the XAFS study was also prepared
using Y$_{2}$O$_{3}$ in the same conditions. The sample of HoBaCo$_{4}$O$_{7}$
was made first in Japan and exactly the same synthesis was reproduced in
Russia. Two samples showed the reproducible values of $T_{\text{s}}$ and
lattice parameters. Other samples, doped with $^{57}$Fe, were prepared in
Russia. Final annealing in flowing pure Ar gas was always done at 600$^{o}$C.

The measurements of x-ray diffraction patterns for Rietveld analysis were
conducted at the "Anomalous Scattering" beamline of the VEPP-3 storage ring in
Siberian Synchrotron Radiation Centre (SSRC). The beamline is equipped with Si
(111) monochromator on the primary beam and Ge (111) crystal analyzer on the
diffracted beam. Three patterns of HoBaCo$_{4}$O$_{7}$ were measured at 300 K
for three wavelengths shown in Fig.2(a). Fourth pattern was measured at 380 K
for $\lambda=$1.5421 \AA . Structure parameters were refined through the
analysis of full-profile x-ray diffraction intensities using FULLPROF
program\cite{DBW,Full}. Anomalous dispersion corrections for the atomic
scattering factors of Ho,Ba and Co were introduced into the input files of
FULLPROF using the Brenann and Cowan data from DispAnoV2 program.%

%TCIMACRO{\FRAME{ftbpFU}{3.1816in}{4.9666in}{0pt}{\Qcb{Three wavelengths used
%in the diffraction experiments shown on the plot of theoretical atomic
%scattering factor corrections $f\prime$ and $f"$\ for\ Co and Ho (a);
%characteristic area of the high-resolution synchrotron x-ray diffraction
%patterns (Rietveld plots) at $\lambda=1.5421$\ \AA \ in the high-temperature
%(b) and low-temperature (c) phases of HoBaCo$_{4}$O$_{7}$. Structure of the
%low-temperature phase was refined at $T=300K$\ using either S.G.
%$Pbn2_{\mathbf{1}}$\ or $Cmc2_{\mathbf{1}}$, one of which ($Pbn2_{\mathbf{1}}%
%$) is shown in (c). Theoretical positions of the permitted reflections are
%shown in (c) for both $Pbn2_{\mathbf{1}}$\ and $Cmc2_{\mathbf{1}}$\ groups.
%Indicated by the arrow reflection $(105)$\ is among very weak peaks
%extinguished for S.G. $Cmc2_{\mathbf{1}}$, but allowed for S.G.
%$Pbn2_{\mathbf{1}}$.\ }}{\Qlb{f2}}{fig_2.eps}%
%width 3.1816in;  height 4.9666in;  depth 0pt;  original-width 4.0524in;
%original-height 6.3706in;  cropleft "0";  croptop "1";  cropright "1";
%cropbottom "0";  filename 'Fig_2.EPS';file-properties "XNPEU";}}}%
%BeginExpansion
\begin{figure}
[ptb]
\begin{center}
\includegraphics[
natheight=6.370600in,
natwidth=4.052400in,
height=4.9666in,
width=3.1816in
]%
{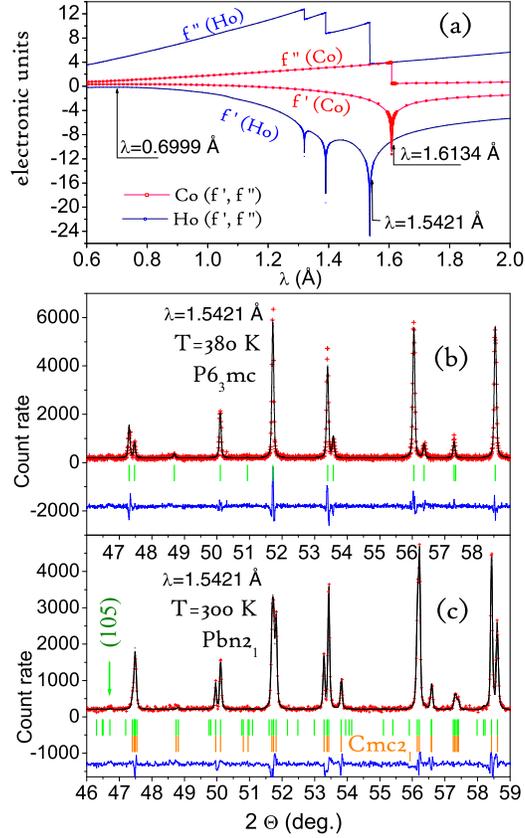}%
\caption{Three wavelengths used in the diffraction experiments shown on the
plot of theoretical atomic scattering factor corrections $f\prime$ and
$f"$\ for\ Co and Ho (a); characteristic area of the high-resolution
synchrotron x-ray diffraction patterns (Rietveld plots) at $\lambda
=1.5421$\ \AA \ in the high-temperature (b) and low-temperature (c) phases of
HoBaCo$_{4}$O$_{7}$. Structure of the low-temperature phase was refined at
$T=300K$\ using either S.G. $Pbn2_{\mathbf{1}}$\ or $Cmc2_{\mathbf{1}}$, one
of which ($Pbn2_{\mathbf{1}}$) is shown in (c). Theoretical positions of the
permitted reflections are shown in (c) for both $Pbn2_{\mathbf{1}}$\ and
$Cmc2_{\mathbf{1}}$\ groups. Indicated by the arrow reflection $(105)$\ is
among very weak peaks extinguished for S.G. $Cmc2_{\mathbf{1}}$, but allowed
for S.G. $Pbn2_{\mathbf{1}}$.\ }%
\label{f2}%
\end{center}
\end{figure}
%EndExpansion

The measurements of x-ray diffraction patterns in the temperature range
between 27$^{o}$C and 113$^{o}$C were also conducted using a MXP21 Mac Science
diffractometer with the following operation conditions: 50$^{o}$
%TCIMACRO{\TEXTsymbol{<} }%
%BeginExpansion
$<$
%EndExpansion
$2\theta$
%TCIMACRO{\TEXTsymbol{<} }%
%BeginExpansion
$<$
%EndExpansion
60$^{o}$ with the step size of 0.02$^{o}$, Cu-K$_{\alpha}$ radiation
($\lambda$= 1.5405 \AA \ and 1.5443 \AA ), V = 45 kV and I = 100 mA. The
lattice parameters were refined using Rietveld analysis.

The Co K-edge XANES spectra were measured at the EXAFS station of SSRC. The
beam was monochromatized with channel-cut Si (111) monochromator. The energy
resolution was 0.8 eV.

M\"{o}ssbauer spectra were collected with the velocity-reversive spectrometer
NZ-640 (Hungary), using the regime of constant acceleration for moving source
of $^{57}$Co embedded in a metal (Cr) matrix. The chemical shifts are given
relative to Fe metal. The temperature of sample was maintained with the
accuracy of $\pm0.1\deg.$ \qquad The M\"{o}ssbauer source was kept at room
temperature. (24$\pm1$ $^{o}$C). In this setup, the drift of the source
temperature had to be taken into account only in the temperature dependencies
of the chemical shift. M\"{o}ssbauer spectra were measured in the range of
velocities from from -4 to 4 mm/s and stored into 1000 channels
equidistributed in velocity. In each of 27 spectra obtained, the number of
counts per channel was approximately $10^{6}$.

\section{Results}

Differential thermal analysis results have revealed the highest temperature of
the structural transition $T_{\text{s}}$ $=355$ K for HoBaCo$_{4}$O$_{7}$
among all the studied previously cobaltites of the RBaCo$_{4}$O$_{7}$ series
(R=Lu,Yb,Tb,Er,Y) (cf. Refs.\cite{Nakayama,Markina,Caignaert}). The DTA curve
collected at the cooling/heating rate of 3 K/min showed the endothermic effect
at the temperature $T_{\text{s}}$ of 82$^{o}$C for sample heating (Fig.3). The
latency interval, within which the high-temperature phase can be undercooled,
evidences the 1-st order character of the transition in agreement with the
general renormalization group theory of the phase transitions (see, for
example, the references in the review \cite{Dove}).%

%TCIMACRO{\FRAME{ftbpFU}{2.0091in}{2.4547in}{0pt}{\Qcb{The curves of
%differential thermal analysis in HoBaCo$_{4}$O$_{7}$ measured upon heating and
%cooling at a rate of 3 deg./min. }}{\Qlb{f3}}{fig_3.eps}%
%width 2.0091in;  height 2.4547in;  depth 0pt;  original-width 5.6224in;
%original-height 6.8862in;  cropleft "0";  croptop "1";  cropright "1";
%cropbottom "0";  filename 'Fig_3.eps';file-properties "XNPEU";}}}%
%BeginExpansion
\begin{figure}
[ptb]
\begin{center}
\includegraphics[
height=2.4547in,
width=2.0091in
]%
{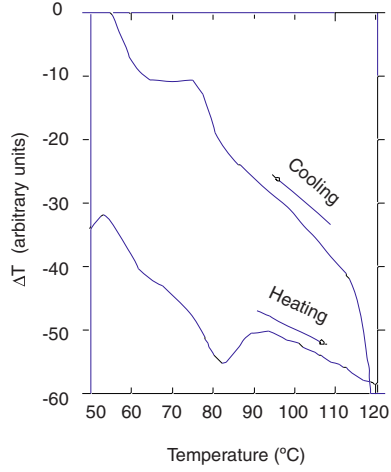}%
\caption{The curves of differential thermal analysis in HoBaCo$_{4}$O$_{7}$
measured upon heating and cooling at a rate of 3 deg./min. }%
\label{f3}%
\end{center}
\end{figure}
%EndExpansion

The spacing between heating and cooling curves is collapsed in Fig. 4, where
the temperature dependence of the lattice parameters is shown. All the data of
Fig. 4 were obtained using the overnight program, so that the average rate of
the temperature "sweep" was slower by two orders of magnitude than that in Fig.3.

\subsection{Structure refinements}

The lattice parameters below and above $T_{s}$ were refined using Rietveld
analysis. To refine the lattice parameters and atomic coordinates at $T<T_{s}
$ we used previously suggested structure models based on the S.G. $Cmc2_{1}$
\cite{Nakayama}, and Pbn2$_{1}$\cite{Huq}. The substitution of 2.5\% Co with
Fe has no significant effect on the lattice parameters. In HoBaCo$_{3.9}%
$Fe$_{0.1}$O$_{7}$ we obtained $a=6.3013(1)$ \AA , b=10.9552(2) \AA ,
c=10.2225(2) \AA . In the Fe-doped sample HoBaCo$_{3.9}$Fe$_{0.1}$O$_{7}$ we
obtained at room temperature the orthorhombic cell with $a=6.3006(3) $ \AA ,
b=10.9573(5) \AA , c=10.2275(5) \AA .%

%TCIMACRO{\FRAME{ftbpFU}{3.0497in}{5.9628in}{0pt}{\Qcb{Evolution of the lattice
%cell parameters $a$, $b/\surd3$ and $c$ with temperature near the structural
%phase transition.}}{\Qlb{f4}}{fig_4.eps}%
%width 3.0497in;  height 5.9628in;  depth 0pt;  original-width 6.2654in;
%original-height 12.3841in;  cropleft "0";  croptop "1";  cropright "1";
%cropbottom "0";  filename 'Fig_4.eps';file-properties "XNPEU";}}}%
%BeginExpansion
\begin{figure}
[ptb]
\begin{center}
\includegraphics[
height=5.9628in,
width=3.0497in
]%
{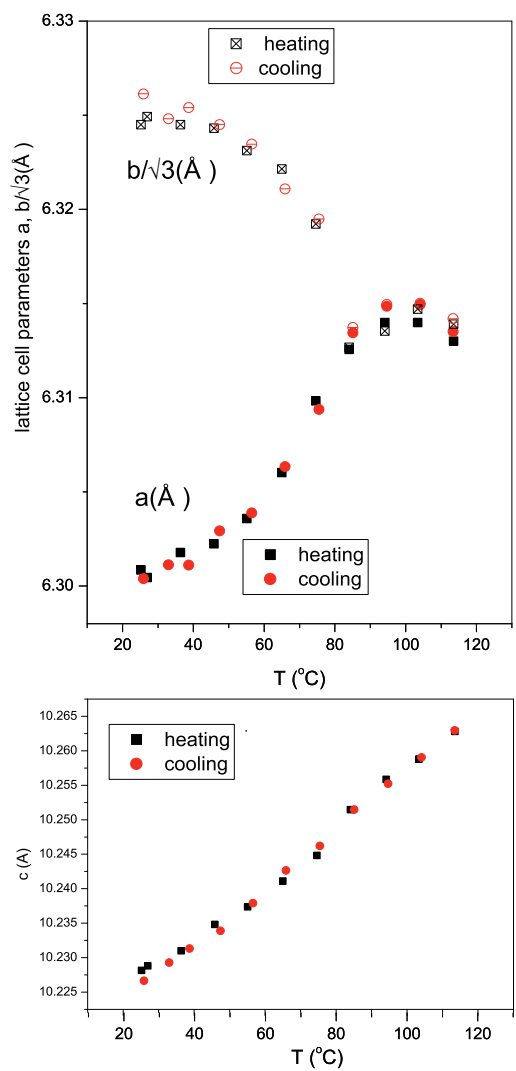}%
\caption{Evolution of the lattice cell parameters $a$, $b/\surd3$ and $c$ with
temperature near the structural phase transition.}%
\label{f4}%
\end{center}
\end{figure}
%EndExpansion

\bigskip

The temperature evolution of the lattice parameter $a$ and the reduced lattice
parameter $b/\sqrt{3}$ shows that there exist below T$_{s}$ a wide temperature
range ($\Delta T=30\div40$ K), in which the orthorhombic distortion gradually
increases at cooling. \ This behavior is contrasting to the abrupt jumping of
the $a$ parameter, reported for YbBaCo$_{4}$O$_{7}$\cite{Maignan}. The
$c$-parameter, on the opposite, exhibits smaller jump at $T_{s}$ in our Fig.4,
as compared to its jump reported previously for YbBaCo$_{4}$O$_{7}%
$\cite{Maignan}.%

%TCIMACRO{\FRAME{ftbpFU}{5.0533in}{3.3126in}{0pt}{\Qcb{The conventional (main
%panel) and differential (inset) Rietveld plots for the synchrotron x-ray
%diffraction data collected at T=300 K (below the phase transition) in
%HoBaCo$_{4}$O$_{7}$. Main panel: observed, calculated and difference
%intensities near Co K-edge ($\lambda=1.6134$ \AA ). Inset: the differences
%$I_{\text{obs}}($CoK$)-$ $I_{\text{obs}}($HoL$_{\text{III}})$ and
%$I_{\text{theor}}($CoK$)-$ $I_{\text{theor}}($HoL$_{\text{III}})$, where the
%profile $I_{\text{theor}}($HoL$_{\text{III}})$ was transformed to the
%$I_{\text{theor}}($CoK$)$ profile conditions.}}{\Qlb{f5}}{fig_5.eps}%
%width 5.0533in;  height 3.3126in;  depth 0pt;  original-width 6.6381in;
%original-height 4.3338in;  cropleft "0";  croptop "1";  cropright "1";
%cropbottom "0";  filename 'Fig_5.EPS';file-properties "XNPEU";}}}%
%BeginExpansion
\begin{figure}
[ptb]
\begin{center}
\includegraphics[
natheight=4.333800in,
natwidth=6.638100in,
height=3.3126in,
width=5.0533in
]%
{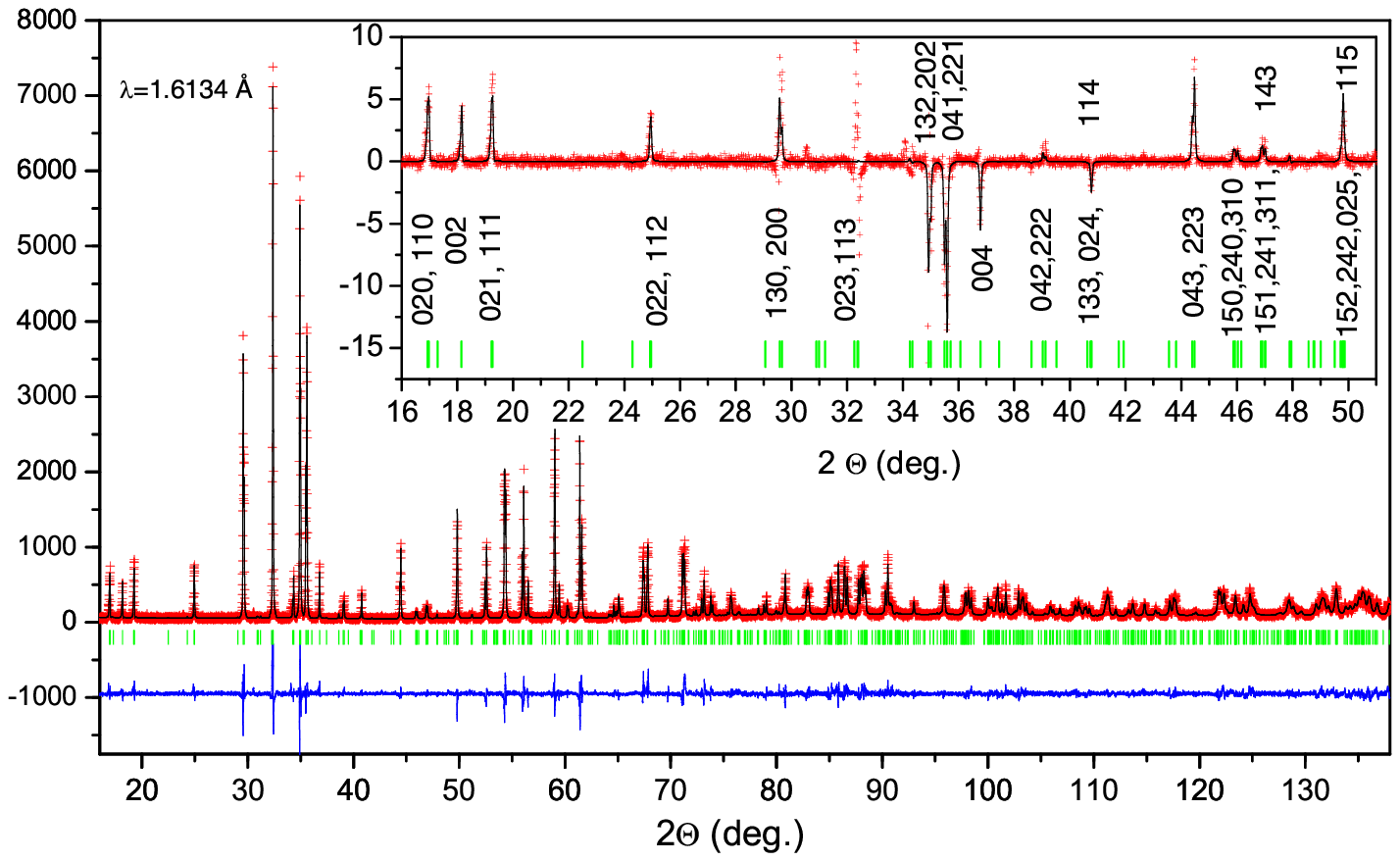}%
\caption{The conventional (main panel) and differential (inset) Rietveld plots
for the synchrotron x-ray diffraction data collected at T=300 K (below the
phase transition) in HoBaCo$_{4}$O$_{7}$. Main panel: observed, calculated and
difference intensities near Co K-edge ($\lambda=1.6134$ \AA ). Inset: the
differences $I_{\text{obs}}($CoK$)-$ $I_{\text{obs}}($HoL$_{\text{III}})$ and
$I_{\text{theor}}($CoK$)-$ $I_{\text{theor}}($HoL$_{\text{III}})$, where the
profile $I_{\text{theor}}($HoL$_{\text{III}})$ was transformed to the
$I_{\text{theor}}($CoK$)$ profile conditions.}%
\label{f5}%
\end{center}
\end{figure}
%EndExpansion

\medskip

\bigskip Table 1. Atomic coordinates for HoBaCo$_{4}$O$_{7}$ at 300 K.\newline%
\begin{tabular}
[c]{|l|ll|ll|ll|ll|}\hline
$Pbn2_{1}$ & x &  & y &  & z &  & $B_{\text{iso}}$ & \\\hline
$Cmc2_{1}$ &  & x &  & y &  & z &  & $B_{\text{iso}}^{\ast}$\\\hline
Ho & {\footnotesize 0.0014(5)} & {\footnotesize 0} & {\footnotesize 0.6672(2)}
& {\footnotesize 0.6664(2)} & {\footnotesize 0.8717(2)} &
{\footnotesize 0.8720(1)} & {\footnotesize 1.4(2)} & {\footnotesize 1.3(2)}\\
Ba & {\footnotesize 0} & {\footnotesize 0} & {\footnotesize 2/3} &
{\footnotesize 2/3} & {\footnotesize 1/2} & {\footnotesize 1/2} &
{\footnotesize 1.4(2)} & {\footnotesize 1.4(2)}\\
Co1 & {\footnotesize -0.011(1)} & {\footnotesize 0} &
{\footnotesize -0.0055(6)} & {\footnotesize 0.0067(4)} &
{\footnotesize 0.9409(4)} & {\footnotesize 0.9388(4)} & {\footnotesize 0.7(2)}
& {\footnotesize 0.9(2)}\\
Co21 & {\footnotesize -0.001(1)} & {\footnotesize 0} &
{\footnotesize 0.1747(2)} & {\footnotesize 0.1770(2)} &
{\footnotesize 0.6996(4)} & {\footnotesize 0.6985(4)} & {\footnotesize 0.6(4)}
& {\footnotesize 0.5(2)}\\
Co22 & {\footnotesize 0.257(1)} & {\footnotesize -} &
{\footnotesize 0.0879(4)} & {\footnotesize -} & {\footnotesize 0.1891(6)} &
{\footnotesize -} & {\footnotesize 1.1(4)} & {\footnotesize -}\\
Co23$^{\ast\ast}$ & {\footnotesize 0.252(1)} & {\footnotesize 0.2525(4)} &
{\footnotesize 0.9239(4)} & {\footnotesize 0.9177(2)} &
{\footnotesize 0.6819(6)} & {\footnotesize 0.6871(4)} & {\footnotesize 1.0(4)}
& {\footnotesize 1.1(3)}\\
O1 & {\footnotesize 0.008(4)} & {\footnotesize 0} & {\footnotesize -0.003(2)}
& {\footnotesize 0.0122(9)} & {\footnotesize 0.2555(7)} &
{\footnotesize 0.2535(7)} & {\footnotesize 1.7(7)} & {\footnotesize 1.0(5)}\\
O21 & {\footnotesize 0.783(2)} & {\footnotesize -} &
{\footnotesize 0.2648(14)} & {\footnotesize -} & {\footnotesize 0.7793(10)} &
{\footnotesize -} & {\footnotesize 0.9(7)} & {\footnotesize -}\\
O22 & {\footnotesize -0.003(5)} & {\footnotesize 0} &
{\footnotesize 0.4860(9)} & {\footnotesize 0.4784(7)} &
{\footnotesize 0.2400(13)} & {\footnotesize 0.2238(7)} &
{\footnotesize 0.6(7)} & {\footnotesize 0.4}\\
O23$^{\ast\ast}$ & {\footnotesize 0.735(2)} & {\footnotesize 0.763(1)} &
{\footnotesize 0.7512(14)} & {\footnotesize 0.7395(7)} &
{\footnotesize 0.2273(12)} & {\footnotesize 0.2600(6)} &
{\footnotesize 0.5(7)} & {\footnotesize 1.6}\\
O31 & {\footnotesize -0.053(2)} & {\footnotesize 0} &
{\footnotesize 0.1535(11)} & {\footnotesize 0.1439(6)} &
{\footnotesize 0.5116(16)} & {\footnotesize 0.5077(9)} &
{\footnotesize 0.4(6)} & $^{\ast\ast\ast}$\\
O32 & {\footnotesize 0.224(2)} & {\footnotesize -} &
{\footnotesize 0.0983(12)} & {\footnotesize -} & {\footnotesize 0.0141(22)} &
{\footnotesize -} & {\footnotesize 1.5(9)} & ${\footnotesize -}$\\
O33$^{\ast\ast}$ & {\footnotesize 0.254(3)} & {\footnotesize 0.250(1)} &
{\footnotesize 0.9355(11)} & {\footnotesize 0.9193(9)} &
{\footnotesize 0.5064(22)} & {\footnotesize 0.5166(8)} &
{\footnotesize 1.5(9)} & $^{\ast\ast\ast\ast}$\\\hline
\end{tabular}

{\footnotesize *Anisotropic thermal displacement factors (ATDF) were refined
for O31 and O33 in S.G. Cmc2}$_{{\footnotesize 1}}${\footnotesize .}

{\footnotesize **Position multiplicity 8 in the S.G. }$Cmc21$%
{\footnotesize \ . All other positions in both }$Pbn2_{1}${\footnotesize \ and
}$Cmc2_{1}${\footnotesize \ groups have the multiplicity 4.}

{\footnotesize ***Refined ATDF for O31: }$U_{{\footnotesize 11}}%
${\footnotesize = 0.13(2), }$U_{{\footnotesize 22}}${\footnotesize =0.016(8),
}$U_{{\footnotesize 33}}${\footnotesize =0.004(9), }$U_{{\footnotesize 23}}%
${\footnotesize =-0.006(8) \AA }$^{2}${\footnotesize , The values of
}$U_{{\footnotesize 13}}$ {\footnotesize and }$U_{{\footnotesize 23}}$
{\footnotesize are fixed to 0 by symmetry. U}$_{{\footnotesize ij}}%
${\footnotesize \ are smaller than B}$_{{\footnotesize ij}}$%
{\footnotesize \ by the factor of 8}$\pi^{2}$.

{\footnotesize ****Refined ATDF for O33: }$U_{{\footnotesize 11}}%
${\footnotesize =0.033(17), }$U_{{\footnotesize 22}}${\footnotesize =0.09(3),
}$U_{{\footnotesize 33}}${\footnotesize =0.028(17), }$U_{{\footnotesize 12}}%
${\footnotesize =0.01(2), }$U_{{\footnotesize 13}}${\footnotesize =-0.01(2),
}$U_{{\footnotesize 23}}${\footnotesize =-0.05(3) \AA }$^{2}${\footnotesize .}

\medskip

Three patterns measured on synchrotron radiation at 300 K were fitted all
together, varying at once the different profile parameters particular of each
pattern and the atomic coordinates common for all patterns. Results shown in
Table 1 were obtained with the S.G.'s $Pbn2_{1}$ and $Cmc2_{1}$. Corresponding
reliability factors are $\chi^{2}=3.83$, $R_{\text{Bragg}}^{1.6134}%
=7.4\%,R_{\text{Bragg}}^{1.5421}=7.7\%,R_{\text{Bragg}}^{0.6999}=9.3\%$ for
$Pbn2_{1}$, and $\chi^{2}=3.91$, $R_{\text{Bragg}}^{1.6134}%
=7.1\%,R_{\text{Bragg}}^{1.5421}=7.2\%,R_{\text{Bragg}}^{0.6999}=9.9\%$ for
$Cmc2_{1}$. The pattern weight factors proportional to the total number of
counts per pattern were used. Quite similar results were found at varying the
weight factor in the broad area of the 3D space of the weight factors.

The use of multiple wavelengths allowed us to vary the contribution of certain
atoms to the diffraction patterns. Diffraction intensities around the angles
of strongest peaks $(2\Theta\sim30-40%
%TCIMACRO{\U{b0}}%
%BeginExpansion
{{}^\circ}%
%EndExpansion
)$ are modified by the anomalous scattering in such a way as if either Ho or
Co were isomorphously replaced with La or Ti, respectively, in effect, as if
the hypothetical compounds LaBaCo$_{4}$O$_{7}$ and HoBaTi$_{4}$O$_{7}$ were
measured additionally to HoBaCo$_{4}$O$_{7}$. Fig.5 shows the Rietveld plot of
HoBaCo$_{4}$O$_{7}$ at the wavelength near Co K-edge. The inset shows the
difference between two resonant patterns measured near the Co and Ho
absorption edges. These differential patterns consist of observed and
theoretical profiles $\Delta I_{\text{obs}}$ and $\Delta I_{\text{theor}}$ at
the profile conditions of the resonant pattern at Co K-edge. To plot the inset
patterns the profiles $I_{\text{obs}}$ and $I_{\text{theor}}$ for
$\lambda=1.5421$ \AA \ were first transformed to the profile conditions of the
profiles $I_{\text{obs}}$ and $I_{\text{theor}}$ for $\lambda=1.6134$
\AA \ and then subtracted from the latter. The difference between experimental
and theoretical differential patterns is largest in a vicinity of the
strongest peak given by the overlapped (023) and (113) reflections. In terms
of both $Pbn2_{1}$ and $Cmc2_{1}$ structural models, the intensity of each of
the (023) and (113) reflections should not vary with simultaneous changing the
Co and Ho scattering power, as is expected from $\Delta I_{\text{theor}}$.
However, it is observed in $\Delta I_{\text{obs}}$ that the (023) reflection
is enhanced at Co K edge, while the (113) reflection is suppressed. The whole
difference profiles including the divergence between $\Delta I_{\text{obs}}$
and $\Delta I_{\text{theor}}$ in the area of (023,113) reflections were
observed to be nearly independent on the choice between the $Pbn2_{1}$ and
$Cmc2_{1}$ structural models.

Among the orthorhombic swedenborgites several compounds were refined
previously with the S.G.'s $Pbn2_{1}$ and $Cmc2_{1}$. The S.G. $Cmc2_{1}$ was
found first to suit the structure of a Zn-based swedenborgite\cite{Rabbow}.
Later, the x-ray diffraction patterns of REBaCo$_{4}$O$_{7}$ for a number of
RE were indexed with the orthorhombic cell of the $Cmc2_{1}$
symmetry\cite{Nakayama}. We observed a few unindexed extra diffraction peaks,
such as the reflection (105) in Fig.2, which could be indexed within the S.G.
$Pbn2_{1}$ suggested recently for YbBaCo$_{4}$O$_{7}$\cite{Huq}. Since the
intensity of the extra reflections is extremely small, their occurrence has,
in fact, a very small effect on the result of refinement. Therefore the
R-factors are quite similar, and even better sometimes for $Cmc2_{1}$ than for
$Pbn2_{1}$. The results for both $Pbn2_{1}$ and $Cmc2_{1}$ symmetries are
shown in Table 1. Resulting structure in terms of $Pbn2_{1}$ is similar to
that of YbBaCo$_{4}$O$_{6.95}$ at 150 K\cite{Huq}. The structure refined in
terms of $Cmc2_{1}$ is shown in Fig. 6.%

%TCIMACRO{\FRAME{ftbpFU}{3.0137in}{2.4501in}{0pt}{\Qcb{Triple unit cell of
%HoBaCo$_{4}$O$_{7}$ refined with the symmetry group $Cmc2_{1}$\ and viewed
%along [001]. The anisotropic thermal displacement factors refined for the
%atoms O31 and O33 are represented by the surface of 50\% probability.}%
%}{\Qlb{f_6}}{fig_6.eps}{\special{ language "Scientific Word";
%type "GRAPHIC";  maintain-aspect-ratio TRUE;  display "USEDEF";
%valid_file "F";  width 3.0137in;  height 2.4501in;  depth 0pt;
%original-width 7.4056in;  original-height 6.0071in;  cropleft "0";
%filename 'Fig_6.eps';file-properties "XNPEU";}}}%
%BeginExpansion
\begin{figure}
[ptb]
\begin{center}
\includegraphics[
height=2.4501in,
width=3.0137in
]%
{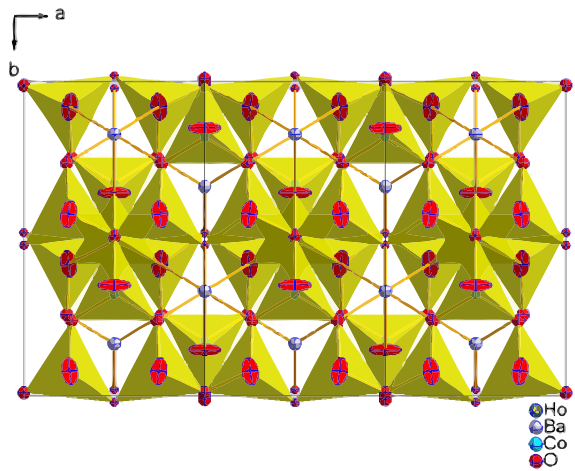}%
\caption{Triple unit cell of HoBaCo$_{4}$O$_{7}$ refined with the symmetry
group $Cmc2_{1}$\ and viewed along [001]. The anisotropic thermal displacement
factors refined for the atoms O31 and O33 are represented by the surface of
50\% probability.}%
\label{f_6}%
\end{center}
\end{figure}
%EndExpansion

In the higher-symmetry phase above $T_{\text{s}}$ there are three
nonequivalent oxygen sites. We call O1 the oxygens in the center of the star
Co$_{4}$O$_{13}$. Other oxygens at the bottom of K-layer in Fig. 1 are called
O2 and those on top of K-layer are called O3. The atoms of cobalt and oxygen
in the Tables 1, 2 and 3 are denoted, depending on the S.G., by single or
double indices to simplify the comparison between the low- and
high-temperature phases. In the high-temperature phase, there occur just two
inequivalent positions for Co labelled as Co1, Co2. With lowering symmetry in
the low-temperature phase, each of the Co2, O2 and O3 splits into two
($Cmc2_{1}$) or three ($Pbn2_{1}$) positions. Then the second index is added
for these positions.

Interatomic distances for the first coordination spheres in the
low-temperature phase are shown in Table 2. Obtained for the S.G. $Pbn2_{1}%
$($Cmc2_{1}$) average distances Co1-O and Co2-O are 1.892(1.900) \AA \ and
1.918 (1.900) \AA , respectively. Average Co-O distances for different Co
sites in the kagom\'{e} layer are $\left\langle d\right\rangle _{\text{Co21-O}%
}=$ 1.921(1.958) \AA , $\left\langle d\right\rangle _{\text{Co22-O}}=$ 1.907
\AA \ and $\left\langle d\right\rangle _{\text{Co23-O}}=$ 1.926(1.871) \AA .
Some preference of the Co1 site for Co$^{3+}$ ions can be suggested only for
the S.G. $Pbn2_{1}$, however, any charge stratification between K and T layers
cannot induce orthorhombicity. Therefore the hypothesis of charge ordering is
implausible within the $Pbn2_{1}$ model. In the $Cmc2_{1}$ model, the state
with Co$^{3+}$ placed into the smallest Co23 site is not a degenerated state
because of double multiplicity of this site compared to the multiplicities of
the remaining Co1 and Co21 sites. Therefore, neither $Pbn2_{1}$ nor $Cmc2_{1}$
models fit the idea of charge ordering.%

%TCIMACRO{\FRAME{ftbpFU}{2.3126in}{4.1917in}{0pt}{\Qcb{[001] projections of the
%crystal structure in high- (top panel) and low- (middle panel) phases in
%HoBaCo$_{4}$O$_{7}$ obtained in this work, and in YbBaCo$_{4}$O$_{7}$ (bottom
%panel, constructed for comparison according to the data of Ref. \cite{Huq}).}%
%}{\Qlb{f_7}}{fig_7.eps}{\special{ language "Scientific Word";
%type "GRAPHIC";  maintain-aspect-ratio TRUE;  display "USEDEF";
%valid_file "F";  width 2.3126in;  height 4.1917in;  depth 0pt;
%original-width 4.057in;  original-height 7.424in;  cropleft "0";
%filename 'Fig_7.eps';file-properties "XNPEU";}}}%
%BeginExpansion
\begin{figure}
[ptb]
\begin{center}
\includegraphics[
height=4.1917in,
width=2.3126in
]%
{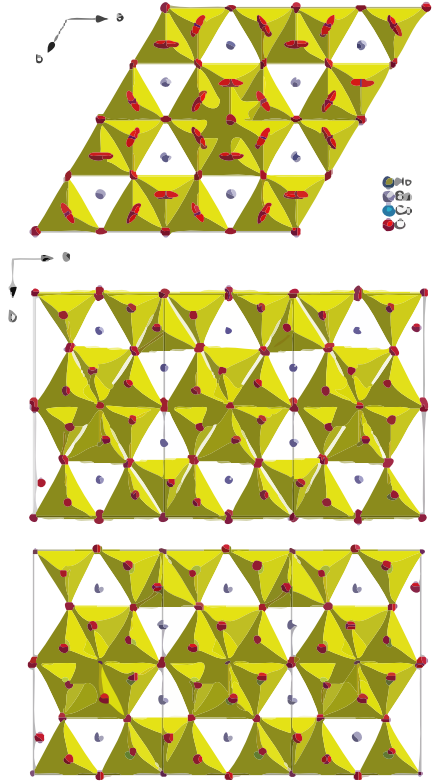}%
\caption{[001] projections of the crystal structure in high- (top panel) and
low- (middle panel) phases in HoBaCo$_{4}$O$_{7}$ obtained in this work, and
in YbBaCo$_{4}$O$_{7}$ (bottom panel, constructed for comparison according to
the data of Ref. \cite{Huq}).}%
\label{f_7}%
\end{center}
\end{figure}
%EndExpansion

We suggest another origin of the orthorhombicity, which is related to locking
the rotational motion of tetrahedra in kagom\'{e} layers. This motion is
visualized with the use of anisotropic thermal factors in Figs. 6 and 7.
Within the refinement of isotropic thermal factors the obtained $B_{\text{iso}%
}$ were too unrealistic ($>4$ \AA $^{2}$) for several oxygen atoms in the
structural models based on the S.G.'s $Cmc2_{1}$ $(T<T_{\text{s}})$ and
$P6_{3}mc$ $(T>T_{\text{s}})$. Therefore, in these models, the anisotropic
motion was allowed at fitting the thermal displacement factors of the O2 and
O3 atoms. This was not necessary for the $Pbn2_{1}$ model $(T<T_{\text{s}}).$
One thus observes at cooling through $T_{\text{s}}$ locking the rotational
motions in the O3 sites only for $Cmc2_{1}$ model, but in both O2 and O3 sites
for $Pbn2_{1}$ model. In the high-temperature phase $P6_{3}mc$, the correlated
motion in the O2 and O3 sites can be thought as a rigid-body libration of the
CoO$_{4}$ tetrahedron. Such a rigid-body picture is not contemplable for the
$Cmc2_{1}$ model, in which only the O31 and O32 sites showed unusual thermal
displacement factors. Related large deformations of tetrahedra are unlikely.
Therefore, through the rest of this article the $Pbn2_{1}$ model of the
low-temperature phase will be considered in more detail. $\bigskip~$

Table 2. Cation-oxygen bondlengths for all Ho, Ba and Co cations in
HoBaCo$_{4}$O$_{7}$ at 300K.\medskip\newline%
\begin{tabular}
[c]{|lll|lll|lll|}\hline
{\footnotesize Bond} & ${\footnotesize Pbn2}_{1}$ & ${\footnotesize Cmc2}_{1}
$ & {\footnotesize Bond} & ${\footnotesize Pbn2}_{1}$ & ${\footnotesize Cmc2}%
_{1}$ & {\footnotesize Bond} & ${\footnotesize Pbn2}_{1}$ &
${\footnotesize Cmc2}_{1}$\\\hline
{\footnotesize Ho-O21} & {\footnotesize 2.291(13)} &  & {\footnotesize Ba-O23}%
$^{^{\ast}}$ & {\footnotesize 3.377(13)} & {\footnotesize 3.298(6)} &
{\footnotesize Co21-O1} & {\footnotesize 1.973(15)} & {\footnotesize 2.144(8)}%
\\
{\footnotesize Ho-O22} & {\footnotesize 2.152(12)} & {\footnotesize 2.190(8)}
& {\footnotesize Ba-O31}$^{^{\ast}}$ & {\footnotesize 2.819(10)} &
{\footnotesize 3.162(5)} & {\footnotesize Co21-O21} &
{\footnotesize 1.867(15)} & \\
{\footnotesize Ho-O23}$^{^{\ast}}$ & {\footnotesize 2.269(14)} &
{\footnotesize 2.264(7)} & {\footnotesize Ba-O31} & {\footnotesize 3.494(10)}
&  & {\footnotesize Co21-O23}$^{^{\ast}}$ & {\footnotesize 1.882(16)} &
{\footnotesize 1.853(7)}\\
{\footnotesize Ho-O31} & {\footnotesize 2.452(14)} & {\footnotesize 2.499(7)}
& {\footnotesize Ba-O32} & {\footnotesize 2.940(14)} &  &
{\footnotesize Co21-O31} & {\footnotesize 1.963(17)} & {\footnotesize 1.98(1)}%
\\
{\footnotesize Ho-O32} & {\footnotesize 2.383(18)} &  & {\footnotesize Ba-O32}
& {\footnotesize 3.387(14)} &  & {\footnotesize Co22-O1} &
{\footnotesize 1.979(15)} & \\
{\footnotesize Ho-O33}$^{^{\ast}}$ & {\footnotesize 2.363(18)} &
{\footnotesize 2.357(9)} & {\footnotesize Ba-O33}$^{^{\ast}}$ &
{\footnotesize 2.970(13)} & {\footnotesize 3.141(1)} &
{\footnotesize Co22-O21} & {\footnotesize 1.866(15)} & \\
&  &  & {\footnotesize Ba-O33}$^{^{\ast}}$ & {\footnotesize 3.353(13)} &
{\footnotesize 3.19(1)} & {\footnotesize Co22-O22} & {\footnotesize 1.978(15)}
& \\\cline{1-6}%
{\footnotesize Ba-O21} & {\footnotesize 2.742(11)} &  &  &  &  &
{\footnotesize Co22-O32} & {\footnotesize 1.805(23)} & \\
{\footnotesize Ba-O21} & {\footnotesize 2.534(12)} &  & {\footnotesize Co1-O1}
& {\footnotesize 1.897(8)} & {\footnotesize 1.910(9)} & {\footnotesize Co23}%
$^{_{-}^{\ast}}${\footnotesize O1} & {\footnotesize 1.996(15)} &
{\footnotesize 1.892(5)}\\
{\footnotesize Ba-O22} & {\footnotesize 2.969(12)} & {\footnotesize 2.785(8)}
& {\footnotesize Co1-O31} & {\footnotesize 1.796(8)} & {\footnotesize 1.80(1)}
& {\footnotesize Co23}$^{_{-}^{\ast}}${\footnotesize O22} &
{\footnotesize 1.927(15)} & {\footnotesize 1.970(5)}\\
{\footnotesize Ba-O22} & {\footnotesize 3.313(13)} & {\footnotesize 3.498(8)}
& {\footnotesize Co1-O32} & {\footnotesize 2.011(17)} &  &
{\footnotesize Co23}$^{_{-}^{\ast}}${\footnotesize O23} &
{\footnotesize 1.977(15)} & {\footnotesize 1.883(8)}\\
{\footnotesize Ba-O23}$^{^{\ast}}$ & {\footnotesize 2.900(13)} &
{\footnotesize 2.979(6)} & {\footnotesize Co1-O33}$^{^{\ast}}$ &
{\footnotesize 1.840(17)} & {\footnotesize 1.945(1)} & {\footnotesize Co23}%
$^{_{-}^{\ast}}${\footnotesize O33} & {\footnotesize 1.802(23)} &
{\footnotesize 1.739(9)}\\\hline
\end{tabular}

{\footnotesize *The indicated bondlengths have double abundance in the S.G.
}${\footnotesize Cmc2}_{{\footnotesize 1}}$ {\footnotesize due to the
multiplicity 8 for the positions of Co23, O23 and O33.}

\medskip{\footnotesize \ }

The lattice parameters refined at $T=$380 K are $a=6.3046(2)$ \AA \ and
$c=10.2604(4)$ \AA . The S.G.'s $P31c$ and $P6_{3}mc$ were examined and gave
the quite similar parameters of fitting quality. In the Table 3, the atomic
coordinates obtained within the S.G. $P6_{3}mc$ are shown. In a recent study
of single crystals of HoBaCo$_{4}$O$_{7}$\cite{Juarez}, the arguments in favor
of the latter S.G. were obtained from observing the equal intensities of
symmetry-equivalent reflections. For example, the reflections (101) and (011)
are symmetry-equivalent. Such distinguishable in single crystals reflections
should have different intensities for $P31c$. In our Rietveld refinements with
both S.G.'s $P6_{3}mc$ and $P31c$, we found while using $B_{\text{iso}}$, that
the S.G. $P31c$ permits to reach a better fit than the S.G. $P6_{3}mc$.
However, with the S.G. $P6_{3}mc$ the equally good fitting quality parameter
($R_{\text{Bragg}}=5.65$ ) was readily obtained through introducing the
anisotropic factors of thermal displacements $B_{ij}$ for O1 and O3 atoms. The
refined anisotropic parameters for these atoms are shown as footnotes under
Table 3.

\medskip

Table 3. Atomic coordinates for HoBaCo$_{4}$O$_{7}$ at 380 K. S.G. $P6_{3}mc$.%

\begin{tabular}
[c]{|l|llll|}\hline
Atom & $x$ & $y$ & $z$ & $B_{\text{iso}}${\footnotesize *}\\\hline
Ho & 2/3 & 1/3 & 0 & 0.4(3)\\
Ba & 2/3 & 1/3 & 0.6258(7) & 1.5(3)\\
Co1 & 0 & 0 & 0.569(3) & 0.4(2)\\
Co2 & 0.1705(7) & -0.1705(7) & 0.805(2) & 0.4(2)\\
O1 & 0 & 0 & 0.357(5) & 1.8(9)\\
O2 & 0.498(3) & -0.498(3) & 0.852(4) & {\footnotesize **}\\
O3 & 0.837(3) & -0.837(3) & 0.131(4) & {\footnotesize ***}\\\hline
\end{tabular}

{\footnotesize *Anisotropic thermal displacement factors were refined for O1
and O3.}

{\footnotesize **}$U_{{\footnotesize 11}}${\footnotesize =}%
$U_{{\footnotesize 22}}${\footnotesize =0.026(9), }$U_{{\footnotesize 33}}%
${\footnotesize =0.07(1), }$U_{{\footnotesize 12}}${\footnotesize =0.02(1),
}$U_{{\footnotesize 13}}${\footnotesize =-}$U_{{\footnotesize 23}}%
${\footnotesize =-0.005(5) \AA }$^{{\footnotesize 2}}${\footnotesize .}

{\footnotesize ***}$U_{{\footnotesize 11}}${\footnotesize =}%
$U_{{\footnotesize 22}}${\footnotesize =0.12(2), }$U_{{\footnotesize 33}}%
${\footnotesize =0.01(1), }$U_{{\footnotesize 12}}${\footnotesize =0.11(2),
}$U_{{\footnotesize 13}}${\footnotesize =-}$U_{{\footnotesize 23}}%
${\footnotesize =-0.002(4).}

\bigskip Each of the Co sites in the kagom\'{e} layer is larger than the Co1
site.\medskip\newline

In the high-temperature phase, we found the average distances Co1-O and Co2-O
of 1.96(2) \AA \ and 1.86(2) \AA , respectively. From low- to high-temperature
phases, we observe, irrespective of the S.G. assumed for the low-temperature
structure, that the average Co2-O distances are shortened, while the average
Co1-O distances are elongated. Especially, for the S.G. $Pbn2_{1}$ each of the
Co sites in the kagom\'{e} layer is larger for $T<T_{\text{s}}$ , but opposite
is true for $T>T_{\text{s}}$. The larger size for the Co1 site, $\left\langle
d\right\rangle _{\text{Co1-O}}>$ $\left\langle d\right\rangle _{\text{Co2-O}}$
is the argument against the larger valence of Co1 than that of Co2. The same
trends were noticed by Huq et al\cite{Huq}, although these authors have
refined the high-temperature structure with the S.G. $P31c$. We note, however,
that the cobaltites REBaCo$_{4}$O$_{7}$, in which a larger valence was found
for Co1 than for Co2 \cite{Schweika}, did not show any structural phase
transition. Judging from the relationship between $\left\langle d\right\rangle
_{\text{Co1-O}}$ and $\left\langle d\right\rangle _{\text{Co2-O}}$ such the
transitionless cobaltites must be "in the low-temperature phase" with vanished
by some reason structural distortion. The observed inversion of this
relationship at heating through $T_{\text{s}}$ has inspired our effort to
search for the changes across $T_{\text{s}}$ in the spectra of XAFS and M\"{o}ssbauer.

\subsection{X-ray absorption spectra near Co K-edge}%

%TCIMACRO{\FRAME{ftbpFU}{2.9648in}{4.1677in}{0pt}{\Qcb{XANES spectra at the Co
%K-edge and the second XANES derivatives in the region of the pre-edge peak for
%HoBaCo$_{4}$O$_{7}$\ (1), YBaCo$_{4}$O$_{7}$\ (2) and YBaCo$_{3.9}$Fe$_{0.1}%
%$O$_{7}$\ (3).}}{\Qlb{f_8}}{fig_8.eps}{\special{ language "Scientific Word";
%type "GRAPHIC";  maintain-aspect-ratio TRUE;  display "USEDEF";
%valid_file "F";  width 2.9648in;  height 4.1677in;  depth 0pt;
%original-width 7.6353in;  original-height 10.7873in;  cropleft "0";
%filename 'Fig_8.eps';file-properties "XNPEU";}}}%
%BeginExpansion
\begin{figure}
[ptb]
\begin{center}
\includegraphics[
height=4.1677in,
width=2.9648in
]%
{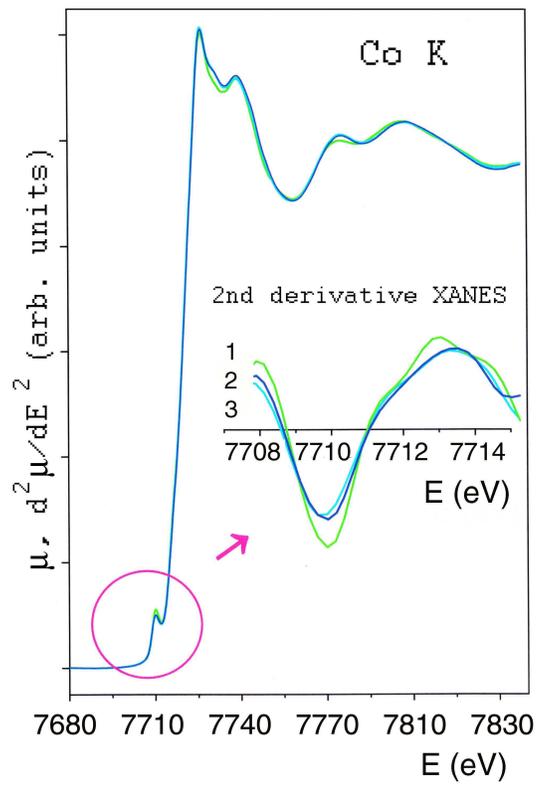}%
\caption{XANES spectra at the Co K-edge and the second XANES derivatives in
the region of the pre-edge peak for HoBaCo$_{4}$O$_{7}$\ (1), YBaCo$_{4}%
$O$_{7}$\ (2) and YBaCo$_{3.9}$Fe$_{0.1}$O$_{7}$\ (3).}%
\label{f_8}%
\end{center}
\end{figure}
%EndExpansion

The XAFS spectra of HoBaCo$_{4}$O$_{7}$ and YBaCo$_{4-x}$Fe$_{x}$O$_{7}$
($x$=0 and 0.1) are very similar (Fig. 8) The XANES begins with a pre-edge
peak at E=7710 eV followed by the main edge and the "white line" maximum of
absorption at 7730 eV. These pre-edge peak and white line correspond to the
electron transitions from 1s orbital to the 3d and 4p bands respectively.
Derivative spectroscopy using the second derivative of XAFS ensures the
effective enhancement of resolution, which can be useful to separate two or
more overlapping subbands. The pre-edge peak in tetrahedrally coordinated Co
site is composed of the vacant orbitals situated in the e$_{g}$ and t$_{2g}$
subbands. Even if they are unresolved in the pre-edge peak as wide as 5 eV,
taking the second derivative of the XAFS spectra could be useful to resolve
them as the separated minima of $\partial^{2}\mu/\partial E^{2}$. In an
octahedral Co site of La$_{1-x}$Sr$_{x}$CoO$_{3}$, for example, such a
manipulation allows to resolve the e$_{g}$ and t$_{2g}$ subbands separated by
2.5 eV \cite{Sikol}.

Because our XAFS station was not equipped with a furnace, all measurements
were done at 300 K. Instead of measuring the spectra of a REBaCo$_{4}$O$_{7}$
across the phase transition, the spectra were measured for RE=Ho and Y with
$T_{\text{s}}$=355 K and 280 K, respectively\cite{OurT}. The only difference
between the spectra for RE=Ho and Y is that the high-energy shoulder of
$\partial^{2}\mu/\partial E^{2}$ near $E$=7712 eV is better articulated in the
low-temperature phase (RE=Ho). In a tetrahedral coordination, the e$_{g}$
subband lies lower in energy than the t$_{2g}$ subband. Therefore, for the
high-spin Co$^{2+}$ ($d^{7}$) the $e_{g}$ orbitals are full and the $t_{2g}$
orbitals are half-filled. The predominant minimum of $\partial^{2}\mu/\partial
E^{2}$ should be ascribed to the $t_{2g}$ orbitals of the high-spin Co$^{2+}$
associated with 75\% of the total pre-edge peak area. Remaining 25\% are due
to Co$^{3+}$. The K-edge Co$^{3+}$ is expected to move to higher energy as a
whole, including the pre-edge peak, by ca. 2 eV. In the high-spin (HS) state
of the tetrahedrally coordinated Co$^{3+}$, the splitting of the pre-edge peak
is expected, since one half-filed orbital appears in the $e_{g}$ subzone,
however, in the intermediate-spin (IS) state of such Co$^{3+}$, the pre-edge
peak is unsplit again as it would come entirely from the transition
$1s\rightarrow3d(t_{2g})$. Therefore, the feature near $E$=7712 eV must be
better pronounced for the lower spin-state of Co$^{3+}$. Comparing Seebeck
coefficient measured in YbBaCo$_{4}$O$_{7}$ with theoretically predicted
spin-dependent values Maignan et al.\cite{Maignan} found that both Co$^{2+}$
and Co$^{3+}$ are in HS states. Our spectra are consistent with the HS states
of Co$^{2+}$ and Co$^{3+}$, although a presence of some IS\ Co$^{3+}$,
especially in low-temperature phase, cannot be excluded.

\subsection{M\"{o}ssbauer spectra}

Our M\"{o}ssbauer spectra collected in the temperature range between ambient
temperature and 133$^{o}$C consist of two series. First, from one measurement
to another, the temperature was ascending. In second series, the measurements
were made for a sequence of temperatures at cooling the sample. All the
obtained spectra are quite symmetric doublets (Fig.9). Spectra treatments are
in order, that would take into account the occurrence of four non-equivalent
sites for the Fe dopants at $T<T_{\text{s}}$ and two sites at $T>T_{\text{s}}%
$. However, up to date it was feasible only to fit each spectrum either with
two-doublet or one-doublet (averaged) envelope. In the two-doublet spectra
treatment, the following parameters of the doublets were obtained at room
temperature: $\delta=0.182$ mm/s; $\varepsilon=$ 0.389 mm/s and $\Gamma=0.342$
mm/s for the majority (80\%) subspectrum, and $\delta=0.193$ mm/s;
$\varepsilon=$ 0.693 mm/s and $\Gamma=0.351$ mm/s for the minority (20\%)
subspectrum.
%TCIMACRO{\FRAME{ftbpFU}{4.1363in}{3.1871in}{0pt}{\Qcb{A typical M\"{o}ssbauer
%spectrum in HoBaCo$_{3.9}$Fe$_{0.1}$O$_{7}$ ($T$= 300 K).}}{\Qlb{f_9}%
%}{fig_9.eps}{\special{ language "Scientific Word";  type "GRAPHIC";
%width 4.1363in;  height 3.1871in;  depth 0pt;  original-width 4.3061in;
%original-height 3.3089in;  cropleft "0";  croptop "1";  cropright "1";
%cropbottom "0";  filename 'Fig_9.eps';file-properties "XNPEU";}}}%
%BeginExpansion
\begin{figure}
[ptb]
\begin{center}
\includegraphics[
height=3.1871in,
width=4.1363in
]%
{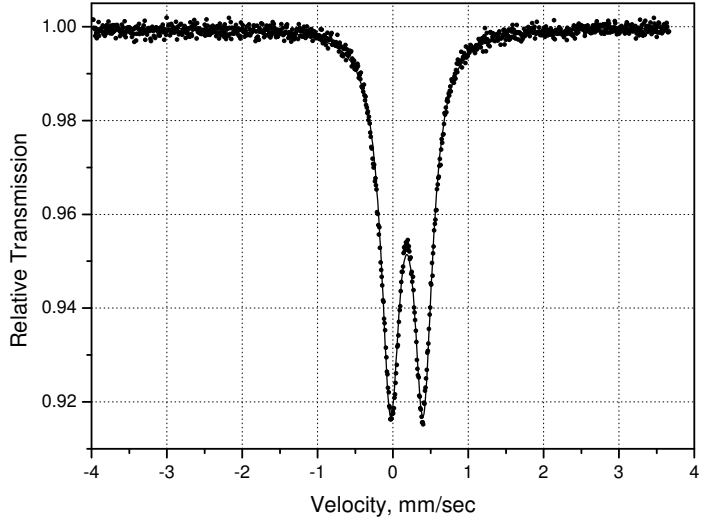}%
\caption{A typical M\"{o}ssbauer spectrum in HoBaCo$_{3.9}$Fe$_{0.1}$O$_{7}$
($T$= 300 K).}%
\label{f_9}%
\end{center}
\end{figure}
%EndExpansion

In the studied range of temperatures 297 K $\leq T\leq$ 406 K, the average
quadrupole splitting is varied by $\simeq$0.01 mm/s only. In such a narrow
range, the determination of the behavior of the spectral parameters was
feasible first of all due to high accuracy of our experimental setup. Second,
a special analysis of the M\"{o}ssbauer data treatments was performed. Namely,
the accuracy of determination of spectral parameters in each spectrum was
taken into account as described below.%

%TCIMACRO{\FRAME{ftbpFU}{4.1659in}{6.1676in}{0pt}{\Qcb{Temperature dependence
%of the chemical shift ($\delta$) and quadrupolar splitting ($\varepsilon$) in
%HoBaCo$_{3.9}^{{}}$Fe$_{0.01}$O$_{7}$. Open triangles shows the data obtained
%point by point sequentially at heating the sample; filled triangles shows the
%$\varepsilon$-data obtained sequentially at cooling the sample. The straight
%line in upper panel is the linear fit to the experimental data:\bigskip
%\ $\delta=A+BT.$}}{\Qlb{f_10}}{fig_10.eps}%
%width 4.1659in;  height 6.1676in;  depth 0pt;  original-width 8.3308in;
%original-height 12.3989in;  cropleft "0";  croptop "1";  cropright "1";
%cropbottom "0";  filename 'Fig_10.eps';file-properties "XNPEU";}}}%
%BeginExpansion
\begin{figure}
[ptb]
\begin{center}
\includegraphics[
height=6.1676in,
width=4.1659in
]%
{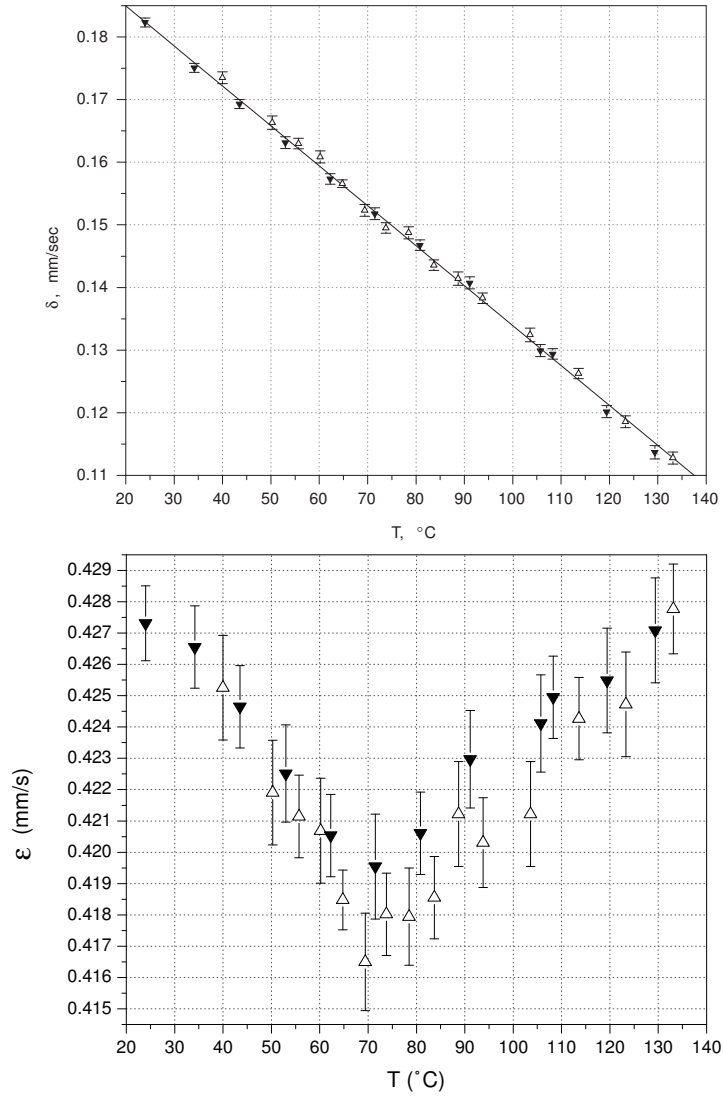}%
\caption{Temperature dependence of the chemical shift ($\delta$) and
quadrupolar splitting ($\varepsilon$) in HoBaCo$_{3.9}^{{}}$Fe$_{0.01}$O$_{7}%
$. Open triangles shows the data obtained point by point sequentially at
heating the sample; filled triangles shows the $\varepsilon$-data obtained
sequentially at cooling the sample. The straight line in upper panel is the
linear fit to the experimental data:\bigskip\ $\delta=A+BT.$}%
\label{f_10}%
\end{center}
\end{figure}
%EndExpansion

Parameters of the M\"{o}ssbauer spectra were determined by a standard
technique via minimizing the value of $\chi^{2}:$%

\begin{equation}
\chi^{2}=\underset{i=1}{\overset{N}{\sum}}\frac{\left[  F_{ex}(i)-F_{th}%
(i;a)\right]  ^{2}}{\sigma_{i}^{2}},\text{ \ \ \ \ \ \ \ }\sigma_{i}%
=\sqrt{F_{ex}(i)}%
\end{equation}
i.e. finding the vector of parameters $a$ of the theoretical spectrum
$F_{th}(i;a)$, such that $\chi^{2}=\min.$ In this work, for all spectra we
adopted the generalized treatment, in which the parameter vector $a$ was
represented by four elements only: (1) average number of counts at infinite
velocity, (2) chemical shift $\delta$, (3) quadrupole splitting $\varepsilon$;
(4) linewidth of the lines of the doublet. Here i=1,...,N are the experimental
points, $F_{ex}(i)$ is the number of counts in the $i^{\text{th}}$ channel and
$\sigma_{i}$ is the evaluation of the dispersion in the experimental spectrum.

During the spectra measurements in this work, the spectrometer operating time
up to 1 month has been reached, however, the errors in determining the
parameters from the minimization of $\chi^{2}$ in each spectrum (Eq.1) do not
involve any drifts and/or uncontrolled instabilities related to a long-term
experimental run. In order to estimate the relative contributions of the
long-term sources of the errors the following procedure was adopted. First, we
assumed that the chemical shift must depend on the temperature linearly as it
does the so-called second-order-Doppler (SOD) shift, or "red shift".
Therefore, we fitted 27 obtained values of the chemical shift with the
straight line. The errors of chemical shift $\Delta_{n}^{2}$ obtained from
non-linear regression of each of these 27 spectra (Eq.(1)) were used in this
linear fit. As a result we obtained an estimation of the squared deviation of
the experimental chemical shift values (i.e., $\delta(n)$ as a component of
the parameter vector $a$ in Eq.(1)) from\ the linear temperature dependence hypothesis:%

\begin{equation}
\chi^{2}(\{\Delta_{n}^{2}\})=\underset{n=1}{\overset{S}{\sum}}\frac{\left[
\delta(n)-A-BT_{n}\right]  ^{2}}{\Delta_{n}^{2}}%
\end{equation}
Here S=27, A=0.198 mm/s and B=-6.38$\cdot10^{-4}$ (mm/s) $\cdot$K$^{-1}$ are
the parameters of linear temperature dependence of the chemical shift,
obtained through minimizing the value of $\chi^{2}$ in the Eq.(2). The value
of A refers to $T$=273 K.

It was supposed then that the uncontrolled long-run instabilities of the
spectrometer can be taken into account through introducing a coefficient
$\eta$ according to the following substitution $\Delta_{n}^{2}\longrightarrow
\eta^{2}\Delta_{n}^{2}$. The value of $\eta$ can be estimated at the condition
that the minimum of $\chi^{2}$ from Eq.(2) reaches its theoretical value when
$\Delta_{n}^{2}$ is replaced with $\eta^{2}\Delta_{n}^{2}$:%

\begin{equation}
\chi^{2}(\{\eta^{2}\Delta_{n}^{2}\})=S-2
\end{equation}

Here $S-2$ is the number of degrees of freedom for the random variables in
Eq.(2). The calculation performed according to this procedure has led to the
result $\eta=1.6.$ Next, the errors in estimations of the chemical shifts and
quadrupole splitting parameters were multiplied by the coefficient $\eta$. The
result of this operation is shown in Figs. (2) and (3). Finally, our analysis
resulted in the average error of the quadrupole splitting not exceeding 0.0017
mm/s. The presented analysis is thus suggesting that the observed variation of
$\varepsilon$ near $T_{\text{s}}$, although small, is the intrinsic and
reproducible property of the material.

\section{Discussion}

The generalized procedure of analyzing the M\"{o}ssbauer spectra allowed us to
treat in the same way the spectra above and below $T_{\text{s}}$=355 K. More
detailed treatments of our spectra, to be presented elsewhere, would take into
account the occurrence of four non-equivalent sites for Co at $T<T_{\text{s}}$
and two sites for Co at $T>T_{\text{s}}$. Recently, two subspectra were
resolved at 4K in YBaCo$_{3.96}$Fe$_{0.04}$O$_{7.02}$ owing to a difference of
magnetic hyperfine fields $\Delta B_{\text{hf}}=3$T\cite{Tsipis}. The authors
claim that the M\"{o}ssbauer spectra cannot resolve the difference between the
sites Co21, Co22 and Co23, however, it is not clear from their article
\cite{Tsipis}, whether or not the orthorhombic distortion was present at all
in their sample. No orthorhombicity down to 100 K was observed recently in
single crystals of HoBaCo$_{4}$O$_{7}$\cite{Juarez}, however, the reason of
such a suppression of the phase transition is yet not understood. When fitting
our M\"{o}ssbauer spectra of HoBaCo$_{4}$Fe$_{0.1}$O$_{7}$ at 300 K (Fig.9)
with two doublets we observe the ratio of the areas of the doublets 4:1, with
the 80\% abundance of the narrower doublet. This is different from the ratio
reported by Tsipis et al. (3:2)\cite{Tsipis}, although the radii of Y and Ho
are similar, so that the distribution of Fe dopants over the Co sites is
expected to coincide. However, the difference of distribution could be related
to the structural difference if the orthorhombicity in the samples of these
authors\cite{Tsipis} is suppressed, similarly to the samples of
Ref.\cite{Juarez}. Another reason for such a divergence may arise as a purely
mathematical trick if the M\"{o}ssbauer spectrometer instrumental linewidths
and lineshapes are very different. These arguments justify our choice in the
present work of the method of spectra treatment, which simplify the comparison
of obtained parameters between the samples with subtle structural differences.

Both temperature regions above and below $T_{\text{s}}$=355 K in our
HoBaCo$_{4}$O$_{7}$ were characterized by very similar M\"{o}ssbauer spectra.
Also, the XAFS spectra of HoBaCo$_{4}$O$_{7}$ and YBaCo$_{4}$O$_{7}$ are very
similar at 300 K despite $T_{\text{s}}$
%TCIMACRO{\TEXTsymbol{>} }%
%BeginExpansion
$>$
%EndExpansion
300 K in HoBaCo$_{4}$O$_{7}$, but $T_{\text{s}}$%
%TCIMACRO{\TEXTsymbol{<} }%
%BeginExpansion
$<$
%EndExpansion
300 K in YBaCo$_{4}$O$_{7}$\cite{OurT}. From these spectra we have shown that
Co$^{2+}$ is in high-spin state. The same conclusion was drawn recently from
the results of soft (Co L$_{2,3}$) x-ray absorption spectroscopy\cite{Hollman}%
. In addition, we have shown that the high-spin state is ubiquitous on both
sides of $T_{\text{s}}$. Concerning the spin state of Co$^{3+}$ the distortion
of tetrahedral coordination would be associated to lifting the degeneracy
either between double degenerated $e_{g}$ states for S=2, or between
triple-degenerated $t_{2g}$ states for S=1. For example, the S=1 spin state
correspond to a distortion which splits off a level for the hole location at
the top of the $t_{2g}$ band. Distortion of a tetrahedron via twisting a pair
of vertices relative to other pair would produce such a splitting\cite{King}.
In both low-and high-temperature phases, we cannot exclude the formation of
intermediate-spin state for Co$^{3+}$. Indeed, obtained distortions for the
CoO$_{4}$ tetrahedra are stronger than the distortions reported in Ref.
\cite{Juarez}.

Observed quadrupole splitting varies within $3\%$ in the range $\pm\Delta
T/T_{\text{s}}$=20\%. The charge redistribution at $T_{\text{s}}$, seen not
only via the change of the Co-O distances, but also via jumps of
resistivity\cite{Caignaert,Maignan}, looks after all very different than a
charge ordering at a metal-insulator transition. Much larger changes of
quadrupole splitting across $T_{\text{s}}$ are usually expected when the
charge ordering is driven by an orbital ordering. Huge jumps of $\varepsilon$
are typically associated with the orbital ordering. For example, in
BiMn$_{0.95}$Fe$_{0.05}$O$_{3}$, the jump-like anomaly $\Delta\varepsilon=0.4$
mm/s of quadrupolar splitting at $T_{\text{OO}}=414K$ was observed\cite{Belik}%
. In the manganites of the family REBaMn$_{1.96}$Fe$_{0.04}$O$_{6}$, fourfold
increase of the quadrupole splitting was observed (at 300K) in the charge and
orbitally ordered compounds (RE=Y,Gd,Sm) compared to unordered ones
(RE=La,Na,Pr) \cite{RJSSC,RJAC,RPRB}. The orbital ordering coupled with charge
ordering is therefore unlikely to drive the structural phase transition at
$T_{\text{s}}$.

Huq et al.\cite{Huq} suggested the origin of the transition to be a "response
to a markedly underbonded Ba$^{2+}$ site in the high-temperature phase". The
ion Ba$^{2+}$ occupies an anticubooctahedral site. In an ideal
anticubooctahedral coordination, there are 12 equidistant ligands. The
geometry of the anticubooctahedron close to ideal one was found in several
studies on single-crystalline samples of REBaCo$_{4}$O$_{7}$ for
RE=Ho\cite{Juarez}, Lu\cite{Kozeeva}, however, severely distorted \ in other
structural studies on single-\cite{VA} and poly-\cite{Huq,Valldor}crystalline
samples. The Ba-O distances of 3.14$\pm0.01$ \AA \ reported for nearly ideal
anticubooctahedron by Juarez-Arellano et al.\cite{Juarez} in HoBaCo$_{4}%
$O$_{7}$ and Kozeeva et al.\cite{Kozeeva} in LuBaCo$_{4}$O$_{7}$ seem to be
too large for ordinary bondlengths of Ba$^{2+}$. No phase transition was
reported in such samples. In other studies, the distortion of the
anticubooctahedron was reported to shorten the shortest Ba-O bondlength down
to 3.08 \AA \ in YBaCo$_{4}$O$_{7}$\cite{VA}, 3.02 \AA \ in HoBaCo$_{4}$%
O$_{7}$\cite{Valldor}, 2.8 \AA \ in YbBaCo$_{4}$O$_{7}$\cite{Huq}. The average
Ba-O distances of 3.14$\pm0.02$ are nearly independent of RE, however, the
distortions in much broader range were reported depending on the symmetry.

In each of the structure models, $P6_{3}mc$, $P31c$, $Cmc2_{1}$ and $Pbn2_{1}
$, the distortion of the anticubooctahedron makes a number of Ba bondlengths
increasing, and the same number of Ba bondlengths decreasing. The average Ba-O
distance remains unchanged. Namely, within the $P6_{3}mc$ model, 6 oxygens of
12 remains at the ideal distances, 3 of them are allowed to move inwards and 3
other outwards of the ideal anticubooctahedron. For the $Cmc2_{1}$ model, such
a ratio changes from 3:6:3 to 5:2:5. The lower symmetries $P31c$ and
$Pbn2_{1}$ allow the distortion ratio 6:0:6, however, in YbBaCo$_{4}$O$_{7}$
Huq et al.\cite{Huq} obtained within the $P31c$ model the anticubooctahedron
distortion scheme 3:6:3 similar to that of $P6_{3}mc$ model. This means the
very small changes of bondlengths of 6 oxygens compared to the ideal
anticubooctahedron. Therefore, the authors argued that the change of the
distortion scheme of the anticubooctahedron from "3:6:3"-like above
$T_{\text{s}}$ to 6:0:6 below $T_{\text{s}}$ is at the origin of the phase
transition. This conclusion is not contradictive to our viewpoint, the more
especially as we fitted successively the high-temperature phase only with the
$P6_{3}mc$ model, which is intrinsically distorted according to the same 3:6:3
scheme. Although both the $P6_{3}mc$ and $P31c$ models result in similar
R-factors for the high-temperature phase, only the $P6_{3}mc$ model gave us
plausible interatomic distances. Indeed, when trying to fit the
high-temperature phase with the $P31c$ model, we obtained the distortion of
the anticubooctahedron much larger than that reported by Huq et al.\cite{Huq}.
In addition, the unrealistic Co-O distances of the order of 1.7 \AA
\ resulted, which obliged us to reject the $P31c$ model.

Thus the driving force of the transition to lower symmetry is likely to arise
from too loose bonding of Ba$^{2+}$ in the symmetric anticubooctahedron. While
the Ba$^{2+}$ coordination polyhedron tends to collapse with lowering
temperature, we do not observe that the coordination polyhedra of the Co ions
change so dramatically. Displacive phase transitions in the tetrahedral
networks are frequently driven by a soft optic phonon, which can propagate
without distortions of tetrahedra. Such modes were called rigid-unit modes
(RUMs)\cite{Boysen}.

The coordination polyhedron of Ba$^{2+}$ is formed by O2 and O3 atoms. Each of
these oxygens enters the first coordination sphere of two Co, two Ba$^{2+} $
and one Ho$^{3+}$. On the other hand, the O1 atom enters to the first
coordination sphere of four Co ions. Therefore, the deformation of the
Ba$^{2+}$ coordination correspond to the motion of O2 and O3 atoms, which
would be also represented in terms of kagom\'{e} network as the motion of the
of the O2 and O3 tetrahedral vertices around the O1 vertex. This is because
each Co in the kagom\'{e} layer (Co2 site) is coordinated by one O1, one O3
and two O2 atoms.

Since the more covalent bonding is stronger within the framework of the
CoO$_{4}$ tetrahedra compared to the ionic bonding on the exterior of the
framework, the distortion of the tetrahedra will cost more energy. The
associated with such a distortion phonon modes would lie rather high on the
frequency scale. On the opposite, the RUMs that do not involve any significant
stretching of the Co-O bonds will have low frequencies. One of the low-energy
RUMs appears to be strongly temperature-dependent to become a soft mode
driving the phase transition. This mode permits the intertetrahedral motions
associated with bending the Co-O-Co links, where the central O atom can be
viewed as a "hinge". The rotations of tetrahedra around the "hinges" are
hindered either by the size of collapsed cavity around Ba$^{2+}$ or by
repulsion between oxygens getting too close to each other. Such hindered
rotations named librations are commonly observed in tetrahedral networks
\cite{Bartelmehs}. In Fig.11, the O-"hinge" can be identified at the O1 atom
(exactly at the center of the top panel in Fig. 7).%

%TCIMACRO{\FRAME{ftbpFU}{2.915in}{2.3993in}{0pt}{\Qcb{The star-like fragment of
%the kagom\'{e} layer showing the surrounding of the O1 "hinge".}}{\Qlb{fig11}%
%}{fig11.eps}{\special{ language "Scientific Word";  type "GRAPHIC";
%width 2.915in;  height 2.3993in;  depth 0pt;  original-width 3.0958in;
%original-height 2.5423in;  cropleft "0";  croptop "1";  cropright "1";
%cropbottom "0";  filename 'Fig11.eps';file-properties "XNPEU";}}}%
%BeginExpansion
\begin{figure}
[ptb]
\begin{center}
\includegraphics[
height=2.3993in,
width=2.915in
]%
{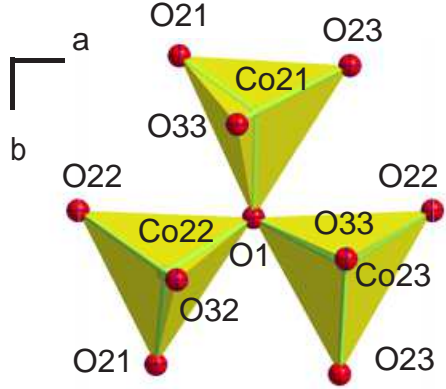}%
\caption{The star-like fragment of the kagom\'{e} layer showing the
surrounding of the O1 "hinge".}%
\label{fig11}%
\end{center}
\end{figure}
%EndExpansion

From Fig. 7 it is evident that our low-temperature phase is quite similar to
that reported for YbBaCo$_{4}$O$_{7}$\cite{Huq}. It is also clear that the
low-temperature phase can be obtained from the high-temperature one via
quenching the positions of the O3 atoms at the endpoints of a counterclockwise
rotation. Since the O3 site splits into three below $T_{\text{s}}$ the
rotation angles are different (in the range between 5 and 10 deg.) for O31,
O32 and O33. The rotation axis [001] is the same for O31, O32 and O33,
therefore, these oxygens keep their planar arrangement as in high-temperature
phase (see the central plane at the bottom of T-layer in Fig.1).

According to the RUMs concept, the axial rotation of the tetrahedral vertices
located on top (in terms of Fig.1) of the kagom\'{e} layers cannot occur
without $z$-displacements of the O2 atoms at the bottom of these kagom\'{e}
layers. This is because only 25\% of the oxygens at the bottom of kagom\'{e}
layers are located on the rotation axes. These O1 atoms constitute the
rotation "hinges". Other 75\% oxygen sites in this plane are the O2 sites,
which split into O21, O22 and O23 in the low-temperature phase. Due to the
rotation of the tetrahedral "rigid units" the O21 and O22 atoms get out of the
plane, in which they were in the high-temperature phase.

The shift of the O21 site along $z$-axis is as large as 0.52 \AA . This is
consistent with the largest rotation angle $(\approx10^{\circ})$ for the O31
atom, which belongs to the tetrahedra of Co21 (in Fig.7 two of these
tetrahedra form a vertical "sandglass" figure extended along $b$-axis around
the central O1 atom). Smaller rotation angles are shown in Fig.7 for the O32
and O33 atoms, which belong to the tetrahedra of Co22 and Co23, respectively.
This is in agreement with smaller out-of-plane $z$-shift of the O22 atoms
(0.14 \AA ), which belong to both Co22 and Co23 tetrahedra. The third site O23
remains at the same $z$-level as in high-temperature phase. This site belongs
to Co21 and Co23 tetrahedra.

Thus, the tetrahedra faces, which were parallel in high-temperature phase,
become slightly tilted relative to horizontal plane. The measure of tilts for
the bottom faces of Co21 and Co23 tetrahedra are the $z$-shifts of
corresponding vertices, i.e., 0.52 \AA \ and 0.14 \AA , for Co21 and Co23
tetrahedra, respectively. The degree of tilt for the Co22 tetrahedra, which is
formed by the O21 and O22 vertices is measured by the difference of their
$z$-shifts, is $0.52-0.14=0.38$ \AA . This relationship directly follows from
the symmetry Pbn2$_{1}$. Because of such a relationship between the values of
$z$-shifts for O21, O22 and O23 atoms the spread between the tilts of three
tetrahedra surrounding the O1-hinge becomes unavoidable. Quite similar spread
was obtained by Huq et al.\cite{Huq}. If the tetrahedra are indeed behaving as
rigid-units, the related spread would in turn appear in the angles of rotation
of the three O3 atoms around the O1---$z$ axis. Comparing the results of our
refinement with the data of Ref. \cite{Huq}, we show in Fig.10 that the spread
is quite similar for HoBaCo$_{4}$O$_{7}$ and YbBaCo$_{4}$O$_{7}$.

In high-temperature phase, where all three O3 atoms around the O1---$z$ axis
are equivalent, they librate with equally large amplitudes as shown at the
upper panel of Fig.7. The thermal displacements of O3 atoms have no component
along $z$, however, the $z-$component is the main component for O2 atoms (see
the footnote of Table 3). However, because the rigidity of tetrahedra put the
constraint on the $z-$shifts of three O2 atoms, the rotations cannot be
condensed below $T_{\text{s}}$ within the high-symmetry model $P6_{3}mc$, in
which all of the O2 sites are equivalent. Therefore, the rigid-body libration
of tetrahedra lowers the symmetry, so that the orthorhombicity directly
follows from combining the RUM condensation with translation symmetry.

In tetrahedral networks, various constraints on possible models of deformation
of these networks were described previously by Dove et al\cite{DGHHP}. In the
anion-close-packed systems, the structure becomes highly susceptible to the
libration-like motions when the proportion of "bridgelike" two-coordinated
anions is large. For example, in the red mercury HgI$_{2}$, where all the
anions are bridgelike, the coupled librations of HgI$_{4}$ tetrahedra are
unconstrainted and the thermal displacements of anions are strongly
anisotropic\cite{Schwarzenbach}. This is not the case for the built of
tetrahedra and octahedra structure of spinel M$_{3}$O$_{4}$, in which all the
oxygens are four-coordinated. A detailed comparizon of the structure of
REBaCo$_{4}$O$_{7}$ with spinels, hexaferrites, perovskites and with
YBaFe$_{4}$O$_{7}$ ferrite was recently made by Caignaert et al\cite{C2009}.
In these structures, the fraction of bridgelike oxygen increases with
increasing the fraction of large-size cations. Like in the red mercury, in the
perovskites, all the oxygens are bridgelike and the perovskites are reach by
their famous soft modes.

Phase transitions known for the tetrahedral and octahedral networks are
classified into displacive transitions and the transitions of the
order-disorder type. In a simple model of double-well potential\cite{Dove},
the type of transition is determined by the relationship between energy of
harmonic forces between neighboring atoms $(\sim k_{\text{B}}T_{\text{s}})$
and the height of the barrier $V_{0}$ between two minima of the potential
well. \ The degree of displaciveness for a transition is given by the
parameter $s^{-1}=$ $k_{\text{B}}T_{\text{s}}/V_{0}$. We are in the displacive
limit when $s^{-1}\gg1$ and in the order-disorder limit when $s\gg1$.
Interestingly, an intermediate type of transition ($s\simeq1$) was attributed
to freezing the librations of large amplitude, such as librations of CO$_{3}$
groups to angle $\sim30^{\circ}$ in calcite\cite{Dove}.

In \ the displacive limit, there exist a well established relationship between
the transition temperature and the angle of rotation of tetrahedra at 0 K:%

\begin{equation}%
%TCIMACRO{\FORMULA{k_{\text{B}}T_{\text{s}}=K\varphi_{0}^{2}}{k_{\text{B}%
%}T_{\text{s}}=K\varphi_{0}^{2}}{evaluate}}%
%BeginExpansion
k_{\text{B}}T_{\text{s}}=K\varphi_{0}^{2}%
%EndExpansion
\label{k-6}%
\end{equation}
\linebreak Markina et al.\cite{Markina} and Juarez-Arellano et
al.\cite{Juarez} have reported nearly linear relationship between the value of
$T_{\text{s}}$ in REBaCo$_{4}$O$_{7}$ and the size of ionic radii of RE. The
temperature $T_{\text{s}}$ increases twice when the size of RE increases by
4\%. The radii of Yb and Ho differ by 3\%. However, in Fig.7 we observe that
the values of $\varphi_{0}$ are very similar for RE=Yb and Ho. The force
constant $K$ in the Eq.(4) is the stiffness of the rigid-body unit. Therefore,
if we are in the displacive limit, where the Eq. (1) is valid, the linear
relationship between $T_{\text{s}}$ and $K$ may signify the increase in the
strength of the Co-O bonds with increasing the ionic radius of RE. This
phenomenon can be regarded as the manifestation of the inductive
effect\cite{Etourneau} in the kagom\'{e} network: Co-O bonding is reinforced
at the expense of RE-O bonding. Indeed, all the O2 and O3 sites (85.7\% of all
oxygen sites) are coordinated by two Co, two Ba, and one RE. The inductive
effect\cite{Etourneau} related to decreased covalence of RE-O bonding with
increasing the size of RE would concern these O2 and O3 sites.

Only remaining 14.3\% oxygen sites (O1-sites) are not bound to RE, as they are
coordinated by four Co. With respect to the covalent Co-O bonding, a
bridge-like (O2 and O3) and tetrahedral (O1) oxygen can be distinguished. The
latter shows more ionic bonding and longer Co-O bonds.

All Co ions are, in their turn, containing in their first coordination sphere
three bridgelike oxygens and one tetrahedral oxygen. Two types of ligands for
each Co presents the plausible argument to explain rather narrow lines of the
symmetric doublet observed in the M\"{o}ssbauer spectra. In spite of the
occurrence of four different sites for Co in the low-temperature phase, the
spectrum shown in Fig. 9 evidences the narrow distribution of EFG over these
sites. If we attribute the origin of EFG to strong difference between the
bridgelike and tetrahedral oxygen ligation, then the EFG can be explained in
terms of point charge model for the first coordination sphere. Then due to the
symmetry in high-temperature phase the EFG on each of the Co sites should be
oriented along the longest Co-O1 bonds for both Co1 and Co2 sites.

The tetrahedra are non-ideal rigid bodies, therefore, in low-temperature phase
each of three tetrahedra in kagom\'{e} layer is distorted in its own manner.
However, we observe only the change of $\varepsilon$ within 3\%. This is an
important observation, which would let us to estimate whether or not the
partial contribution of each ligand to EFG depend on the bond angles.
Generally, upon variation of bond angles the partial contribution to EFG\ may
change because fractional $p-$character of the bonds changes\cite{ButzHyp2}.
However, the families of compounds exist, in which the assumption of
angle-independent partial contributions is a good approximation. One of such
families includes, for example, brownmillerites and high-$T_{\text{c}}$
superconductors with "1212" structure\cite{RCR}. The high-$T_{\text{c}}$
cuprates, so different at first sight from brownmillerites, show very similar
ligation for iron located at the midway between the cuprate
planes\cite{RCR,PB,RNUV}.

The quadrupole splitting is related to the\ EFG tensor main component $V_{zz}
$ and EFG asymmetry $\eta=(V_{xx}-V_{yy})/V_{zz}$ via\cite{RNUV}:%

\begin{equation}
\varepsilon=\frac{eQcV_{zz}}{2E_{\gamma}}\left(  1+\frac{\eta^{2}}{3}\right)
^{%
%TCIMACRO{\U{bd}}%
%BeginExpansion
\frac12
%EndExpansion
}.
\end{equation}
Here $E_{\gamma}=14.4125$ keV is the transition energy, and $Q$ is the
quadrupole moment of the excited state of the $^{57}$Fe nuclei. From the
electric field gradient V$_{zz}$ in units of V/m$^{2}$ the quadrupole
splitting in mm/s can be obtained using the factor $%
%TCIMACRO{\U{bd}}%
%BeginExpansion
\frac12
%EndExpansion
\cdot eQcE_{\gamma}^{-1}=0.1664\cdot10^{-21}$mm$\cdot$s$^{-1}$V$^{-1}$m$^{2}$
for the $Q$ value of 0.16 barn. The Eq.(5) shows that the changes in
$\varepsilon$ signifies either variation of $V_{zz}$ or variation of $\eta$ or both.%

%TCIMACRO{\FRAME{ftbpFU}{3.9122in}{6.3161in}{0pt}{\Qcb{(a) Czjzek-plot of
%$Y=V_{\text{zz}}(1-\eta)=-2V_{\text{xx}}$ versus $X=\mid V_{\text{zz}}%
%\mid(\checkmark3+\eta/\checkmark3)$ $=2/\checkmark3\mid2V_{\text{zz}}%
%+V_{xx}\mid$ with the EFG tensor components ordered according to $\mid
%V_{xx}\mid\leq\mid V_{\text{yy}}\mid\leq\mid V_{\text{zz}}\mid$. Lines
%originating from the origin are $\eta=$ const. lines. The angles of
%inclination of the straight-line trajectories with respect to abscissa axis
%are indicated. Arrows show the direction of increasing temperature. (b)
%Corresponding dependences of the EFG asymmetry $\eta(T)$. The frames mark the
%$\eta(T)-$curves for two trajectories whose EFG components are shown in detail
%in Fig. 13. }}{\Qlb{f12}}{fig_12.eps}{\special{ language "Scientific Word";
%type "GRAPHIC";  maintain-aspect-ratio TRUE;  display "USEDEF";
%valid_file "F";  width 3.9122in;  height 6.3161in;  depth 0pt;
%original-width 3.8282in;  original-height 6.223in;  cropleft "0";
%filename 'Fig_12.EPS';file-properties "XNPEU";}}}%
%BeginExpansion
\begin{figure}
[ptb]
\begin{center}
\includegraphics[
natheight=6.223000in,
natwidth=3.828200in,
height=6.3161in,
width=3.9122in
]%
{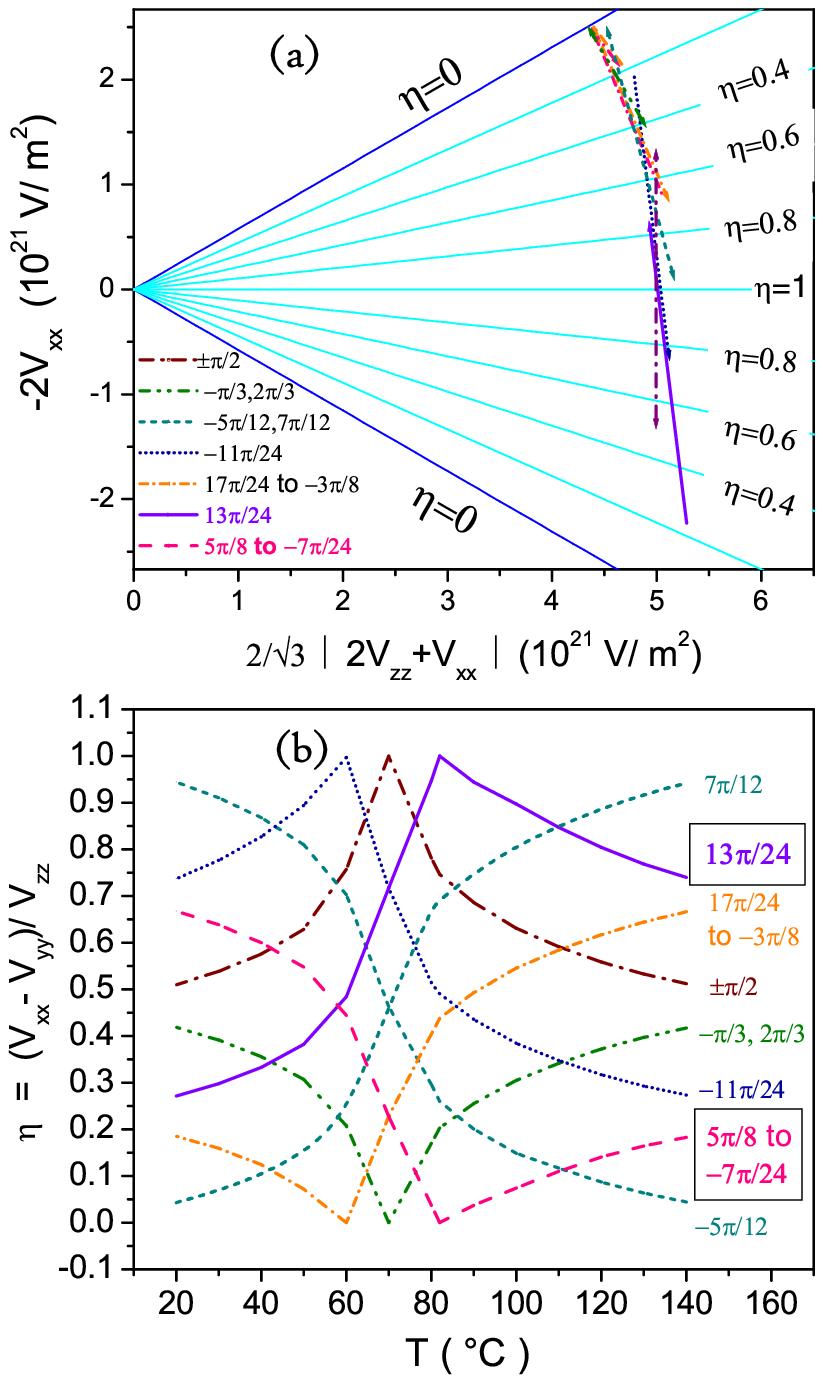}%
\caption{(a) Czjzek-plot of $Y=V_{\text{zz}}(1-\eta)=-2V_{\text{xx}}$ versus
$X=\mid V_{\text{zz}}\mid(\checkmark3+\eta/\checkmark3)$ $=2/\checkmark
3\mid2V_{\text{zz}}+V_{xx}\mid$ with the EFG tensor components ordered
according to $\mid V_{xx}\mid\leq\mid V_{\text{yy}}\mid\leq\mid V_{\text{zz}%
}\mid$. Lines originating from the origin are $\eta=$ const. lines. The angles
of inclination of the straight-line trajectories with respect to abscissa axis
are indicated. Arrows show the direction of increasing temperature. (b)
Corresponding dependences of the EFG asymmetry $\eta(T)$. The frames mark the
$\eta(T)-$curves for two trajectories whose EFG components are shown in detail
in Fig. 13. }%
\label{f12}%
\end{center}
\end{figure}
%EndExpansion

The behavior of $\eta$ is well predictable in a vicinity of a second-order
phase transition. Displacive phase transitions are usually close to the
continuous phase transitions, which are described though rigorously with
renormalization group theory as first-order-like, in practice exhibit the size
of first-order discontinuities so small as to be virtually
unobservable\cite{Dove}. Through the continuous phase transitions an empiric
rule was established\cite{Butz}, which states that all three components of EFG
tensor depends linearly on a single control parameter. In other words, the
changes of principal component and asymmetry of EFG tensor are correlated and
both $V_{zz}$ and $\eta$ vary continuously through such a transition. Symmetry
breaking obviously underlies this rule in a vicinity of axiality ($\eta=0$) or
antiaxiality ($\eta=1$).%

%TCIMACRO{\FRAME{ftbpFU}{3.3144in}{5.0468in}{0pt}{\Qcb{Variation of the EFG
%tensor components for two straight-line billiard trajectories in Czjzek-plot
%with the direction of increasing temperature inclined at the angle to abscissa
%axis of 5$\pi/8$ and reflected to $-7\pi/24$ (a) and $13\pi/24$ (b).
%Corresponding dependences $\eta(T)$ are marked by frames in Fig.12 (b). }%
%}{\Qlb{f13}}{fig_13.eps}{\special{ language "Scientific Word";
%type "GRAPHIC";  maintain-aspect-ratio TRUE;  display "USEDEF";
%valid_file "F";  width 3.3144in;  height 5.0468in;  depth 0pt;
%original-width 4.1409in;  original-height 6.341in;  cropleft "0";
%filename 'Fig_13.EPS';file-properties "XNPEU";}}}%
%BeginExpansion
\begin{figure}
[ptb]
\begin{center}
\includegraphics[
natheight=6.341000in,
natwidth=4.140900in,
height=5.0468in,
width=3.3144in
]%
{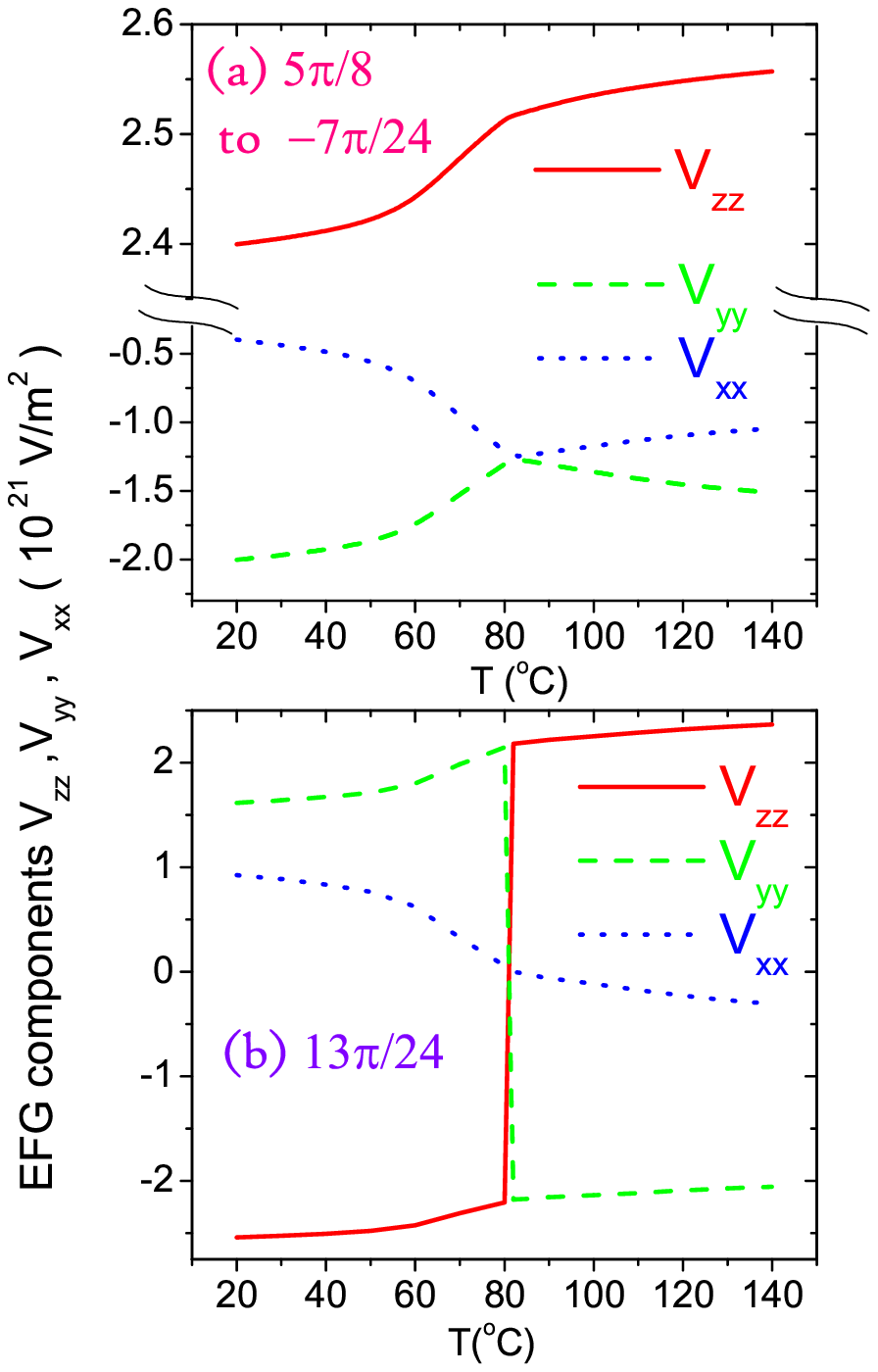}%
\caption{Variation of the EFG tensor components for two straight-line billiard
trajectories in Czjzek-plot with the direction of increasing temperature
inclined at the angle to abscissa axis of 5$\pi/8$ and reflected to $-7\pi/24$
(a) and $13\pi/24$ (b). Corresponding dependences $\eta(T)$ are marked by
frames in Fig.12 (b). }%
\label{f13}%
\end{center}
\end{figure}
%EndExpansion

The non-trivial correlation between $V_{zz}$ and $\eta$ is illustrated by the
linear trajectories of the system in the so-called Czjzek-plot, in which
$-2V_{\text{xx}}$ is plotted versus $\mid V_{\text{zz}}\mid(\checkmark
3+\eta/\checkmark3)$\cite{ButzHyp2,Butz}. These axes are chosen to make use
the intrinsic properties of the EFG tensor, which is symmetric and traceless.
Therefore at the condition of antiaxiality (at horizontal axis in Czjzek plot,
Fig.12a) , where $V_{zz}$ changes its sign by definition, the system pass the
line $\eta=1$ without refraction. On the other hand, at $\eta=0$ the system
reflects from the borders of Czjzek plot with billiard kinematics. These
properties are satisfied always when each component of the EFG tensor depends
linearly on a single control parameter $p$\cite{Butz}. This parameter is in
turn a function of temperature $p=\digamma(T)$.

In Fig.12(b), we plotted $\eta$ versus $T$ assuming several linear paths of
the system in Chjzek-plot shown in Fig. 12(a). First of all, the path at the
angle of $\pi/2$ to horizontal axis crosses the line $\eta=1$ at the same
temperature where $\varepsilon(T)$ shows the minimum in Fig. 10. Therefore,
$\eta(T)$ curve culminates at $T=70^{\circ}$C in Fig.12(b). Also, the
trajectory, which starts at the angle of 2$\pi/3$ near $\eta=0.4$, then goes
to the upper border of the Chjzek-plot, reflects from the $\eta=0$ line
backwards to the angle of -$\pi/3$ and returns back to $\eta=0.4$, produces
the curve $\eta(T)$ with the minimum at $T=70^{\circ}$C. The path tilted to
the midway angle with the directions towards (-$5\pi/12)$ and (7$\pi/12$)
produces two $\eta(T)$ curves with no extremum, but with sigmoidal shape
having the inflexion again at $T=70^{\circ}$C. These two curves correspond to
two opposite directions of increasing temperature along the same line
indicated by arrows on Czjzek-plot (Fig.12,a).

Next, we investigate the trajectory tilted to the angles intermediate between
$7\pi/12$ and 2$\pi/3.$ As in the case of 2$\pi/3,$ such a trajectory is
reflected from the border of Czjzek-plot, but now not at the right angle. The
path starting at the angle of $17\pi/24$ reflects to the angle of $-3\pi/8$.
Moving in the opposite direction along the same trajectory we start from short
segment at the angle of 5$\pi/8$ and reflect to -7$\pi/24.$ In both these
cases, the reflection point does not coincide with the minimum of the
$\varepsilon(T)$ curve in Fig.10. Because our experimental data for
$\varepsilon(T)$ are more or less symmetric relative to the minimum at
$T=70^{\circ}$C, two possible trajectories for these angles generate two
curves for $\eta(T)$. One of them shows minimum at $T=82 $ $^{\circ}$C and
another shows minimum at $T=60$ $^{\circ}$C.

In the same way, two trajectories in the Czjzek-plot and two $\eta(T)$ curves
can be generated for the angles of -$11\pi/24$ and 13$\pi/24$ along the line
intermediate between $7\pi/12$ and $\pi/2.$ These two paths cross the
horizontal axis ($\eta=1$ line) at different temperatures. Again the generated
$\eta(T)$ curves culminate at $T=82$ $^{\circ}$C and at $T=60$ $^{\circ}$C.

Since we know now both values of $\eta(T)$ and $\mid V_{\text{zz}}(T)\mid$
along each path in Czjzek plot, we can derive each component of the EFG
tensor. In the displacive limit, the behavior of the $\eta(T)$ in a vicinity
of $T_{\text{s}}$ is expected to follow closely the behavior of order
parameter around the point where symmetry breaks\cite{Butz}. Two of the
obtained $\eta(T)$ curves culminate indeed at $T_{\text{s}}=82$ $^{\circ}$C.
Therefore, if we are indeed not far from the displacive limit, the EFG
components $V_{\text{zz}}$, $V_{\text{yy}}$ and $V_{\text{xx}}$ should behave
as shown in Fig.13. In the first case (5$\pi/8$ to -7$\pi/24$), the EFG is
nearly axial above $T_{\text{s}}=82$ $^{\circ}$C, but quickly approaches to
antiaxilality as the hexagonal symmetry breaks down. The non-principal EFG
components exchange their direction at $T_{\text{s}}$. The second case
(13$\pi/24$) is opposite. In the second case, the principal EFG component
switch its direction at $T_{\text{s}}$. The plot in Fig.13(a) corresponds to
the appearance of the additional source of EFG below $T_{\text{s}}$, which
induces second large EFG component. On the opposite, the plot in Fig.13(b)
correspond to the case when two large components of EFG exist above
$T_{\text{s}}$ and one of them is extinguished at the expense of the
displacements showing up with the breakdown of symmetry.

Only the first case fits the high-temperature axiality which should correspond
to our classification of the oxygen ligands as bridgelike and tetrahedral
ones. In this case the $z$-axis of the EFG in its internal coordinate system
must be directed from each Co ion towards ionic tetrahedral oxygen O1 (in the
radial direction of Fig. 11). In this picture, the small asymmetry of EFG
above $T_{\text{s}}$ arises from the structural difference between bridgelike
oxygens O2 and O3. Because of non-ideal rigidity of tetrahedra below
$T_{\text{s}}$ the displacements of the O3 atoms are larger than the
displacements of the O2 atoms. The change in the bond angles and the
associated change in the fractional $p-$character of these oxygen bonds result
in the additional component of EFG, which becomes comparable in magnitude with
the main EFG component.

\section{Concluding Remarks}

We have studied the temperature dependence of structure parameters and
measured the hyperfine splitting $\varepsilon(T)$ of $^{57}$Fe in
HoBaCo$_{3.9}$Fe$_{0.1}$O$_{7}$ in the range $\pm\Delta T/T_{\text{s}}$=20\%
around the structural phase transition. Although the changes of $\varepsilon
(T_{\text{s}}\pm\Delta T)$ within only 3\% were observed, we showed that the
non-principal EFG components ($V_{xx}$ and $V_{yy}$) vary much more
dramatically in a vicinity of $T_{\text{s}}=355$ K in agreement with the local
structure changes at the symmetry-breaking transition. Although the variations
of each of three individual EFG components $V_{xx}$, $V_{yy}$ and $V_{zz}$ are
smooth similarly to the temperature dependences of the lattice parameters $a$
and $b/\surd3$, we have demonstrated the nonmonotonic behavior of the
combination $\eta(T)$ = $(V_{xx}-V_{yy})/V_{zz}$, which experiences a dip-like
(or cusp-like) anomaly at $T_{\text{s}}$. In some vicinity of $T_{\text{s}}$
the temperature dependence of the EFG asymmetry $\eta(T)$ is that of the order
parameter as expected for a displacive phase transition.

Rietveld refinements of the structure on both sides of $T_{\text{s}}$ allowed
identifying the main component of thermal motion which freezes as the
hexagonal symmetry breaks down. The origin of the transition is ascribed to
the condensation of the libration phonon mode associated with the rigid-body
rotational movements of starlike tetrahedral units, the building blocks of
kagom\'{e} network.

\section{Acknowledgements}

Authors thank Dan'kova V.S. for the preparation of some of the samples used in
this work. This study was supported by RFBR under Grant No. 07-02-91201 and by
a Grant-in-Aid for Scientific Research (JSPS Japan - RFBR Russia Joint
Research Project and No. 19052004) from the Japan Society for the Promotion of Science.

\section{Figure Captions}

Fig. 1. Perspective view of the quadruple unit cell in the low-temperature
phase of ReBaCo$_{4}$O$_{7}$. Kagom\'{e} and triangular layers of CoO$_{4}$
tetrahedra are marked by "K" and "T". The quadruple unit cell of HoBaCo$_{4}%
$O$_{7}$ is shown with atomic coordinates refined in S.G. $Pbn2_{1}$.

Fig. 2. Three wavelengths used in the diffraction experiments shown on the
plot of theoretical atomic scattering factor corrections $f\prime$ and
$f"$\ for\ Co and Ho (a); characteristic area of the high-resolution
synchrotron x-ray diffraction patterns (Rietveld plots) at $\lambda
=1.5421$\ \AA \ in the high-temperature (b) and low-temperature (c) phases of
HoBaCo$_{4}$O$_{7} $. Structure of the low-temperature phase was refined at
$T=300K$\ using either S.G. $Pbn2_{\mathbf{1}}$\ or $Cmc2_{\mathbf{1}}$, one
of which ($Pbn2_{\mathbf{1}}$) is shown in (c). Theoretical positions of the
permitted reflections are shown in (c) for both $Pbn2_{\mathbf{1}}$\ and
$Cmc2_{\mathbf{1}}$\ groups. Indicated by the arrow reflection $(105)$\ is
among very weak peaks extinguished for S.G. $Cmc2_{\mathbf{1}}$, but allowed
for S.G. $Pbn2_{\mathbf{1}}$.

Fig. 3. The curves of differential thermal analysis in HoBaCo$_{4}$O$_{7}$
measured upon heating and cooling at a rate of 3 deg./min.

Fig. 4. Evolution of the lattice cell parameters a, b/$\surd3$ and c with
temperature near the structural phase transition.

Fig. 5. The conventional (main panel) and differential (inset) Rietveld plots
for the synchrotron x-ray diffraction data collected at T=300 K (below the
phase transition) in HoBaCo$_{4}$O$_{7}$. Main panel: observed, calculated and
difference intensities near Co K-edge ($\lambda=1.6134$ \AA ). Inset: the
differences $I_{\text{obs}}($CoK$)-$ $I_{\text{obs}}($HoL$_{\text{III}})$ and
$I_{\text{theor}}($CoK$)-$ $I_{\text{theor}}($HoL$_{\text{III}})$, where the
profile $I_{\text{theor}}($HoL$_{\text{III}})$ was transformed to the
$I_{\text{theor}}($CoK$)$ profile conditions.

Fig. 6. Triple unit cell of HoBaCo$_{4}$O$_{7}$ refined with the symmetry
group $Cmc2_{1}$\ and viewed along [001]. The anisotropic thermal displacement
factors refined for the atoms O31 and O33 are represented by the surface of
50\% probability.

Fig. 7. [001] projections of the crystal structure in high- (top panel) and
low- (middle panel) phases in HoBaCo$_{4}$O$_{7}$ obtained in this work, and
in YbBaCo$_{4}$O$_{7}$(bottom panel, constructed for comparison according to
the data of Ref. \cite{Huq}).

Fig. 8. XANES spectra at the Co K-edge and the second XANES derivatives in the
region of the pre-edge peak for HoBaCo$_{4}$O$_{7}$\ (1), YBaCo$_{4}$O$_{7}%
$\ (2) and YBaCo$_{3.9}$Fe$_{0.1}$O$_{7}$\ (3).

Fig. 9. A typical M\"{o}ssbauer spectrum in HoBaCo$_{3.9}$Fe$_{0.1}$O$_{7}$
($T$= 300 K).

Fig. 10. Temperature dependence of the chemical shift ($\delta$) and
quadrupolar splitting ($\varepsilon$) in HoBaCo$_{3.9}^{{}}$Fe$_{0.01}$O$_{7}%
$. Open triangles shows the data obtained point by point sequentially at
heating the sample; filled triangles shows the data obtained sequentially at
cooling the sample. The straight line in upper panel is the linear fit to the
experimental data $\delta=A+BT$.

Fig. 11. The star-like fragment of the kagom\'{e} layer showing the
surrounding of the O1 "hinge".

Fig. 12. (a) Czjzek-plot of $Y=V_{\text{zz}}(1-\eta)=-2V_{\text{xx}}$ versus
$X=\mid V_{\text{zz}}\mid(\checkmark3+\eta/\checkmark3)$ $=2/\checkmark
3\mid2V_{\text{zz}}+V_{xx}\mid$ with the EFG tensor components ordered
according to $\mid V_{xx}\mid\leq\mid V_{\text{yy}}\mid\leq\mid V_{\text{zz}%
}\mid$. Lines originating from the origin are $\eta=$ const. lines. The angles
of inclination of the straight-line trajectories with respect to abscissa axis
are indicated. Arrows show the direction of increasing temperature. (b)
Corresponding dependences of the EFG asymmetry $\eta(T)$. The frames mark the
$\eta(T)-$curves for two trajectories whose EFG components are shown in detail
in Fig. 13.

Fig. 13. Variation of the EFG tensor components for two straight-line billiard
trajectories in Czjzek-plot with the direction of increasing temperature
inclined at the angle to abscissa axis of 5$\pi/8$ and reflected to $-7\pi/24$
(a) and $13\pi/24$ (b). Corresponding dependences $\eta(T)$ are marked by
frames in Fig.12 (b).

\bigskip

\begin{thebibliography}{99}
\bibitem {OurT}The $T_{\text{s}}$ value for our YBaCo$_{4}$O$_{7}$ was lower
by 20 K than $T_{\text{s}}$ in YBaCo$_{4}$O$_{7}$ synthesized by Caignaert et
al.\cite{Caignaert}, probably, owing to slightly different conditions of final annealing.
\end{thebibliography}
\end{document}